\documentclass[12pt]{article}
\usepackage{amsmath,amssymb,amscd}
\def\thickone{{\rm 1\mskip-4.5mu l}}
\usepackage[colorlinks=true,linkcolor=blue,urlcolor=blue]{hyperref}
\usepackage[T1]{fontenc}
\def\specialdigits{}
\def\tempcite#1{\hbox{[\quad]}}

\newcommand{\cjs}{c_j(\sigma)}
\newcommand{\hj}{\mskip4mu\hat{\mskip-3mu\text{\it\j}}\mskip2mu}
\newcommand{\bhj}{\mskip4mu\bar{\hat{\mskip-3mu\text{\it\j}}}\mskip2mu}
\newcommand{\fjs}{f_j(\sigma)}
\newcommand{\fj}{f_j} %implied (\sigma)
\newcommand{\fls}{f_{\ell}(\sigma)}
\newcommand{\hatjovfs}{\mathchoice
{\tfrac{\hj }{f_j(\sigma)}}
{\frac{\hj }{f_j(\sigma)}}
{\frac{\hj }{f_j(\sigma)}}
{\tfrac{\hj }{f_j(\sigma)}}}
\newcommand{\hatlovfs}{\mathchoice
{\tfrac{\hat{\ell}}{f_\ell(\sigma)}}
{\frac{\hat{\ell}}{f_\ell(\sigma)}}
{\frac{\hat{\ell}}{f_\ell(\sigma)}}
{\frac{\hat{\ell}}{f_\ell(\sigma)}}}
\newcommand{\nrovrhos}{\mathchoice
{\tfrac{n(r)}{\rho(\sigma)}}
{\frac{n(r)}{\rho(\sigma)}}
{\frac{n(r)}{\rho(\sigma)}}
{\frac{n(r)}{\rho(\sigma)}}}
\newcommand{\nsovrhos}{\mathchoice
{\tfrac{n(s)}{\rho(\sigma)}}
{\frac{n(s)}{\rho(\sigma)}}
{\frac{n(s)}{\rho(\sigma)}}
{\frac{n(s)}{\rho(\sigma)}}}
\newcommand{\hatlovfjs}{\mathchoice
{\tfrac{\hat{\ell}}{f_j(\sigma)}}
{\frac{\hat{\ell}}{f_j(\sigma)}}
{\frac{\hat{\ell}}{f_j(\sigma)}}
{\frac{\hat{\ell}}{f_j(\sigma)}}}
\newcommand{\kd}[2]{\delta_{#1,#2}}
\newcommand{\ket}[1]{\mathop{{|}\mskip1mu #1\mskip1mu\rangle}\nolimits}
\newcommand{\kets}[2]{\mathop{{|}\mskip1mu #1\mskip1mu\rangle}\nolimits\mskip-4mu_{#2}}
\newcommand{\Z}{\mathbb{Z}}
\newcommand{\prefrac}[2]{\tfrac1{#2}{#1}}
\newcommand{\no}[1]{\:{:}\mskip1mu #1\mskip1mu {:}\:}
\newcommand{\nosub}[2]{\:{:}\mskip1mu #1\mskip1mu {:}_{#2}}
\def\cramped#1{\radical0 {\kern-\nulldelimiterspace#1}}
\newcommand{\ps}{(\sigma)}

\def\mfrac#1#2{{\textstyle{\textstyle\medmuskip1mu\hbox{$#1$}\lower2.5pt\hbox{}\over
      \textstyle\medmuskip1mu\hbox{$\cramped{#2}$}\vbox to9pt{}}}}
\newcommand{\sS}[2]{_{#1}^{\smash{\phantom{#1}}#2}}
\let\tsty\textstyle
\newcommand{\sqfjs}{\sqrt{\smash[b]{\fjs}\mskip-2mu}\mskip3mu}
\newcommand{\sqfls}{\sqrt{\smash[b]{\fls}\mskip-2mu}\mskip3mu}
\newcommand{\twone}{e^{2\pi i}}
\def\abs#1{\lvert #1\rvert}
\def\babs#1{\bigl| #1\bigr|}
\def\norm#1{\lVert #1\rVert}
\def\bnorm#1{\bigl\| #1\bigr\|}
\def\BIgl({\mathopen{\hbox{\smaller${\biggl(}$}}}
\def\BIgr){\mathclose{\hbox{\smaller${\biggr)}$}}}
\def\bil({\mathopen{\raise.25pt\hbox{\smaller${\bigl(}$}}}
\def\bir){\mathclose{\raise.25pt\hbox{\smaller${\bigr)}$}}}
\def\unfrac#1#2{#1/#2}
\numberwithin{equation}{section}

\makeatletter
\newenvironment{tcases}{%
  \matrix@check\tcases\env@tcases
}{%
  \crcr\noalign{\vskip-2pt}\endarray\right.%
}
\def\env@tcases{%
  \let\@ifnextchar\new@ifnextchar
  \left\lbrace
  \def\arraystretch{1.0}%
  \array{@{}l@{\quad}l@{}}%
  \noalign{\vskip-4pt}
}
\makeatother

\title{The orbifold-string theories of permutation-type:\\ 
II. Cycle dynamics and target space-time dimensions}
\author{M.~B. Halpern\thanks{halpern@physics.berkeley.edu}\\
Department of Physics\\
University of California, 
Berkeley, California 94720, USA}
\date{}

\begin{document}

\maketitle

\begin{abstract}
We continue our discussion of the general bosonic prototype of the new
orbifold-string theories of permutation-type. 
Supplementing the extended physical-state
conditions of the previous paper, we construct here
the extended Virasoro generators with cycle central charge $\hat{c}_j(\sigma) = 26\fjs$,
where $\fjs$ is the length of cycle~$j$ in twisted sector~$\sigma$. We
also find an equivalent, reduced formulation of each 
physical-state problem at reduced cycle central charge $\cjs=26$.
These tools are used to begin the study of the target space-time dimension $\hat{D}_j(\sigma)$ of cycle~$j$
in sector~$\sigma$, which is naturally defined as the number of
zero modes (momenta) of each cycle. The general model-dependent formulae derived here will
be used extensively in succeeding papers, but 
are evaluated in this paper only for the simplest case
of the ``pure'' permutation orbifolds.
\end{abstract}
\nocite{*}

\clearpage
\tableofcontents

\clearpage

\section{Introduction}\label{Section 1}
% MANU 1.1

The new orbifold-string theories of permutation-type [1-6] 
include the bosonic prototypes
\begin{subequations}
\label{1.1.group1}
\begin{gather}
\label{1.1.1}
\frac{{\rm U}(1)^{26 K}}{H_+} = \frac{{\rm U}(1)_1^{26}\times\dots\times{\rm U}(1)_K^{26}}{H_+},
\quad
H_+\subset H({\rm perm})_K \times H_{26}',
\\
\label{1.1.2}
\left[\frac{{\rm U}(1)^{26K}}{H_+}\right]_{{\rm open}} ,
\\
\label{1.1.3}
\frac{{\rm U}(1)^{26 K}}{H_-} = \frac{{\rm U}(1)_L^{26}\times{\rm U}(1)_R^{26}}{H_-},
\quad
H_{-}\subset \Z_2({\rm w.s.}) \times H'_{26}
\end{gather}
\end{subequations}
and generalizations of these, as noted in 
Appendix B of Ref.~6. The three families in \eqref{1.1.group1} are
called respectively the generalized permutation orbifolds 
(twisted closed strings at sector central charge $\hat{c} = 26 K$),
the open-string analogues of the generalized permutation orbifolds
(twisted open strings at $\hat{c} = 26 K$),
and the orientation-orbifold string systems, which 
contain an equal number of twisted closed strings at $\hat{c} =26$ 
and twisted open strings at $\hat{c} = 52$. The open-string sectors
of the orientation orbifolds are contained, along with their
$T$-duals, at $K=2$ in the open-string analogues 
of the generalized permutation orbifolds. The closed-string sectors of 
the orientation orbifolds form the ordinary space-time orbifold 
$U(1)^{26}/H'_{26}$ at $\hat{c}=26$. Further information
on special cases of these orbifold-string systems, especially $\hat{c} = 52$, is contained in Refs.~[3-5].
We note in particular that the orientation-orbifold string systems (1.1c)
generalize and include [4] the ordinary critical bosonic open-closed string system.

%MANU 1.2
In the previous paper [6] of the present series, cycle-bases of general 
permutation groups and the principles of the orbifold program [7-21] were used to construct a twisted BRST system
for each cycle~$j$ in each twisted sector~$\sigma$ of these 
orbifolds, including the extended algebra of the BRST charges
\begin{equation}
\label{1.2.1}
[ \hat{Q}_i( \sigma ), \hat{Q}_j ( \sigma )]_+ = 0 \quad \hbox{ $\forall i,j$ in sector~$\sigma$},
\end{equation}
and right-mover copies of these systems in the
twisted closed-string sectors. Moreover, the BRST
systems were used to find the \emph {extended physical-state
conditions} of the matter in cycle~$j$ of sector~$\sigma$
%MANU 1.3 
\begin{subequations}
\label{1.3.group1}
\begin{gather}
\label{1.3.1}
\bigl( \hat{L}_{\hj j} ((m+\hatjovfs) \geq 0) - \hat{a}_{\fjs} \kd{m+\hatjovfs}{0}\bigr)
\kets{\chi( \sigma )}{j} = 0, \\
\label{1.3.2}
\begin{gathered}[b]
\bigl[ 
\hat{L}_{\hj j} (m+\hatjovfs) ,
\hat{L}_{\hat{\ell} \ell} (m+\hatlovfs)
\bigr]
\hfill\\ 
=
\delta_{j\ell} 
\bigl\{
(m-n-\tfrac{\hj -\hat{\ell}}{\fjs}) \hat{L}_{\hj+\hat{\ell},j}(m+n+\tfrac{\hj+\hat{\ell}}{\fjs})
\\
\hskip9em\quad{}
+
\prefrac{\hat{c}_j(\sigma)}{12}(m+\hatjovfs)((m+\hatjovfs)^2 - 1)\kd{m+n+\frac{\hj+\hat{\ell}}{\fjs}}{0}
\bigr\},
\end{gathered}
\\
\hat{c}_j(\sigma) = 26 \fjs, \quad 
\hat{a}_{\fjs} = \frac{13 f^2_j( \sigma ) - 1}{12 \fjs},
\\
\bhj = 0,1,\dots,\fjs -1, \quad
j = 0,1,\dots, N( \sigma ) - 1, \quad
\sum_j \fjs = K
\end{gather}
\end{subequations}
%MANU 1.4
including again a right-mover copy of these conditions for twisted 
closed-string sectors. The algebra (1.3b) of the matter generators $\{\hat{L}_{\hj j} \}$ is called the 
\emph{orbifold Virasoro algebra} [7,15,6] of cycle $j$ in sector~$\sigma$.
The fundamental numbers (1.3c) of each cycle are the \emph{cycle
central charge} $\hat{c}_j( \sigma )$ and the \emph{cycle-intercept} $\hat{a}_{\fjs}$,
both expressed in terms of the length $\fjs$ of cycle~$j$ in sector~$\sigma$.
Using the final sum rule in Eq.~(1.3d) the reader easily 
verifies that the \emph{sector central charges}
\begin{equation}
\hat{c}( \sigma ) = \sum_j \hat{c}_j(\sigma)
\end{equation}
are $26K$ for the closed- and open-string counterparts
of the generalized permutation orbifolds and $52 \,(K = f_0( \sigma ) = 2)$
for the twisted open-string sectors of the orientation orbifolds.
The twisted closed-string sectors of the orientation orbifolds (the ordinary
space-time orbifold $U(1)^{26}/H'_{26}$)
can also be obtained from these results by choosing
$K= N(\sigma)=f_{0}(\sigma) =1$ and hence the ordinary values $\hat{a}_1 = 1, \hat{c}(\sigma)=\hat{c}_{0}(\sigma)=26$. 

%%MANU 1.5
In fact these results see 
only the permutation subgroup $H ({\rm perm})_K$ or $\Z_2 ({\rm w.s.})$
of $H_\pm$,
which determines the twisted permutation gravities [2]
of each sector and hence the BRST systems. The 26-dimensional
automorphism subgroup $H'_{26}$ of $H_\pm$, which operates
uniformly on each left- and right-mover copy of the
critical closed string, is encoded however in the explicit form
of the extended Virasoro generators of the matter.
  
%is not encountered until we
%ask for the explicit 
%form of the orbifold Virasoro generators as a function of the twisted 
%matter currents.

%MANU 1.6
Our first task in this paper is therefore  to
supplement the extended physical-state conditions (1.3)
with the construction of the orbifold Virasoro
generators $\{ \hat{L}_{\hj j}\}$ at cycle central charge  $\hat{c}_j(\sigma) =\nobreak 26 \fjs$
 as functions of the twisted matter.
%for any choice of the 26-dimensional automorphism subgroup $H'_{26}$. 
% at $\hat{c}_j(\sigma) =\nobreak 26 \fjs$
%in cycle~$j$ of sector~$\sigma$.
This construction encodes the solution of the spectral problem
of each element $\omega( \sigma)  \in H'_{26}$ of the 26-dimensional
automorphism subgroup, and our general formulae will be evaluated 
explicitly for a large class of examples  of $H'_{26}$ in the
following paper. Subexamples of this construction at $\hat{c}(\sigma)=52$ 
and $\hat{c}(\sigma)=26\lambda, \lambda$ prime have already been discussed in Refs.~ [3-5].
%The general forms given here
%are the proper venue for the consideration of
%the broad features of 
%the entire class of orbifold-string theories.

%%MANU 1.7
Generalizing our work at $\hat{c}(\sigma)=52$ in Ref.~[3], we shall also find an
equivalent, \emph {reduced} form of the physical-state problem for each
cycle~$j$ of each sector $\sigma$ at \emph {reduced} cycle central charge $\cjs = 26$.

%%MANU 1.8
With these tools , we shall begin a survey of
%At both the unreduced and the reduced level,
%we shall begin to consider 
the \emph{space-time $($target-space$)$ interpretation}
of the orbifold-string theories, noting in particular with Ref.~[3] that the 
target space-times are invariant
under the reduction.
% as discussed for $\hat{c}( \sigma ) = 52$
%in Refs.~\tempcite{}. 
In this discussion, we will focus on the
\emph{target space-time dimension} 
\begin{equation}
\hat{D}_j(\sigma) \equiv \dim \{ \hat{J}_j(0)_{ \sigma } \}
\end{equation}
of cycle~$j$ in sector~$\sigma$  as the number of
zero modes (momenta) of the
cycle, and following Refs.~[3-5], we will define the momentum-squared
operators and level-spacing which are needed to analyze 
the extended physical-state problems. In the only explicit examples
of this paper, we shall find that $\hat{D}_j(\sigma) = 26$ for
the ``pure'' permutation orbifolds (with trivial $H'_{26}$), so that these simple cases
are equivalent to collections of ordinary critical strings. More generally
however \emph {the dimensionality of the target space-time is not equal to any of the 
central charges of the theories}, and in succeeding papers we will present many examples
of non-trivial $H'_{26}$ with $\hat{D}_{j}(\sigma)\leq 26$!

%In large
%examples of non-trivial $H'_{26}$  we shall see however
%in succeeding papers that the dimensions
%of these space-times are in general \emph{not} equal
%to the central
%charge $\hat{c}_j\ps = 26 \fjs$, 
%nor the reduced central charge $\cjs = 26$, but can in fact be less than 26!

\section{An application of the orbifold program}\label{Section 2}
%MANU 2.1
To obtain the general forms of the 
orbifold Virasoro generators for each
cycle~$j$ of each sector~$\sigma$, we apply the standard
methods of the orbifold program [7-21] which emphasizes
the principle of local isomorphisms [9,11,12,15-17].

%MANU 2.2
The orbifold program always
begins with the operator-product formulation of the untwisted
systems in question. Here we need then only
the operator-product form of the stress-tensor/current
system of $K$ copies of the critical bosonic string:
%%MANU 2.3
\begin{subequations}
\label{2.3.group1}
\begin{gather}
\label{2.3.1}
T_I(z) = \tfrac12 G^{ab} \no{ J_{aI}(z) J_{bI}(z)},
\quad
I = 0,1,\dots,K-1,
\\
\label{2.3.2}
G = 
\Bigl(
\hskip-2pt
\begin{array}{rc}
-1 & 0 \\
%[-7pt]
 0 & \thickone
 %\mathbb{1}
\end{array}
\Bigr)
,
\quad
a,b = 0,1,\dots,25,
\\
\label{2.3.3}
\begin{gathered}[b]
T_I(z) T_J(w) 
= \delta_{IJ} \biggl( \frac{26/2}{(z-w)^4} + \Bigl(\mfrac{2}{(z-w)^2} + \mfrac{1}{z-w} \partial_w\Bigr)
T_I(w)\biggr)\quad{}\\
\hfill{}
 +
\no{ T_I(z) T_J(w) }
\end{gathered}
\\
\label{2.3.4}
\begin{gathered}[b]
T_I(z) J_{aJ}(w) 
= \delta_{IJ} \Bigl( \mfrac{1}{(z-w)^2} + \mfrac{1}{z-w} \partial_w\Bigr)J_{aI}(w)
\qquad{}
\\
\hfill{} +
\no {T_I(z) J_{aJ}(w) }
\end{gathered}
\\
\label{2.3.5}
J_{aI} (z) J_{bJ} (w) = 
\frac{\delta_{IJ} G_{ab}}{(z-w)^2} +
\no{J_{aI} (z) J_{bJ} (w) }.
\end{gather}
\end{subequations}
%%MANU 2.3'
The symbol $\no{\cdots}$ is operator-product normal 
ordering, that is, the operator product minus the
singularities shown. The total stress tensor $T(z) = \sum_I T_{I}(z) $
is also Virasoro with total central charge $c = 26K$, and a right-mover
copy of this system is included implicitly in
our application below.

%%MANU 2.4
The automorphic responses of these 
operators are
\begin{subequations}
\label{2.4.group1}
\begin{gather}
\label{2.4.1}
J_{aI} (z)' = \omega( \sigma) \sS{a}{b} W( \sigma )\sS{I}{J} J_{bJ}(z), \\
\label{2.4.2}
T_{I}(z)' = W( \sigma )\sS{I}{J} T_{J}(z), \\
\label{2.4.3}
W( \sigma ) \in H ({\rm perm} )_K, \quad
\omega( \sigma)  \in H'_{26}.
\end{gather}
\end{subequations}
Note that the definition of a sector~$\sigma$ requires the choice
of one element $(W\times \omega)$ from each equivalence class of both $H({\rm perm})_K$ and $H'_{26}$.
%MANU 2.5
Here we are following the sector-labeling convention of the orbifold program for 
product groups, 
but we mention that $\sigma$ can equivalently be viewed as a two-component 
vector, with one component each for $H({\rm perm})_K$ and $H'_{26}$.

The next step in the orbifold program is to find the so-called 
eigenfields [9,11,12,15-17] under the
automorphism groups, and for this we must first recall the \emph{$H$-eigenvalue problems} of
the group elements: For each element $W( \sigma )\in H({\rm perm})_K$ 
we have the spectral problem
\begin{subequations}
\label{2.5.group1}
\begin{gather}
\label{2.5.1}
W( \sigma )\sS{I}{J} V^\dagger ( \sigma )\sS{J}{\hj j} = 
V^\dagger( \sigma )\sS{I}{\hj j} e^{-2\pi i \frac{\hj}{\fjs}},
\\
\label{2.5.2}
\bhj = 0,1,\dots, \fjs - 1, \quad
j = 0,1,\dots, N( \sigma ) - 1,
\\
\label{2.5.3}
{\textstyle\sum\limits_j} = N( \sigma ), \quad
{\textstyle\sum\limits_j} \fjs = K,
\end{gather}
\end{subequations}
where $j$ labels cycles of length $\fjs$, $N( \sigma )$
is the number of cycles in $W( \sigma )$, and $\hj$ indexes within each cycle~$j$.
The explicit form of the unitary eigenmatrix $V( \sigma )$ is given in Refs.~[13.15], and 
this eigenvalue problem was also discussed in the previous paper [6]
on the general BRST problem.  What is essential to add here is 
the eigenvalue problem for each element $\omega( \sigma)  \in H'_{26}$ of the
26-dimensional automorphism subgroup:
\begin{subequations}
\label{2.5.group2}
\begin{gather}
\label{2.5.4}
\omega( \sigma) \sS{a}{b} U^\dagger( \sigma )\sS{b}{n(r)\mu} = 
U^\dagger ( \sigma )\sS{a}{n(r)\mu} e^{-2\pi i \nrovrhos }
\\
\label{2.5.5}
\bar{n}(r)\in \{0,1,\dots,\rho(\sigma) - 1\},\quad
{\tsty\sum\limits_\mu} = \dim [\bar{n} (r)], \quad
{\tsty\sum\limits_r} \dim [\bar{n} (r)] = 26.
\end{gather}
\end{subequations}
Here $U(\sigma)$ is the unitary eigenmatrix of $\omega(\sigma)$, with 
order $\rho(\sigma)$ ,and $n(r)$, $\mu =\mu(n(r))$ are respectively
the spectral and degeneracy indices of $\omega( \sigma) $.
%and $U(\sigma)$ is the unitary eigenmatrix.
The barred quantities in (2.3b) and (2.4b) are the pullbacks 
of the spectral indices to their fundamental ranges. Many
of these spectral problems [9,11,13,15,16] have been solved explicitly
in the orbifold program , but we will not choose
any particular non-trivial $H'_{26}$ in the general discussion of this
paper (see however Sec.~10).

%%MANU 2.6
Given the forms of these two eigenvalue problems, we may
write down the \emph {eigenfields} for each $W( \sigma )\in H({\rm perm})_{K}$ and $\omega( \sigma) \in H'_{26}$:
\begin{subequations}
\label{2.6.group1}
\begin{gather}
\label{2.6.1}
\varTheta_{\hj j}(z,\sigma)  \equiv 
\sqfjs V( \sigma )\sS{\hj j}{I} T_I(z),\\
\label{2.6.2}
\mathcal{J}_{n(r)\mu\hj j}(z,\sigma)  \equiv
\chi_{n(r)\mu}( \sigma )\sqfjs U( \sigma )\sS{n(r)\mu}{a}
V( \sigma )\sS{\hj j}{I} J_{aI}(z).
\end{gather}
\end{subequations}
Here we have chosen the standard normalization $\chi_{\hj j}( \sigma ) = \sqfjs$
for elements of $H({\rm perm})_K$, 
but left the normalizations $\chi_{n(r)\mu}( \sigma )$ arbitrary for elements of $H'_{26}$.
The eigenfields are constructed to diagonalize the
automorphic responses as follows:
\begin{subequations}
\label{2.6.group2}
\begin{gather}
\label{2.6.3}
\varTheta_{\hj j}(z,\sigma) ' = e^{-2\pi i \frac{\hj}{\fjs}} 
\varTheta_{\hj j}(z,\sigma) ,
\\
\label{2.6.4}
\mathcal{J}_{n(r)\mu \hj j}(z,\sigma) ' = e^{-2\pi i (\frac{\hj}{\fjs} + \nrovrhos )}
\mathcal{J}_{n(r)\mu \hj j}(z,\sigma) .
\end{gather}
\end{subequations}
Moreover, the eigenfields inherit the following periodicity conditions
%%MANU 2.7
\begin{subequations}
\label{2.7.group1}
\begin{gather}
\label{2.7.1}
\varTheta_{\hj \pm \fjs, j} (z,\sigma)  = \theta_{\hj j}(z,\sigma) ,
\\
\label{2.7.2}
\mathcal{J}_{n(r) \pm \rho(\sigma), \mu \hj j}(z,\sigma)  = 
\mathcal{J}_{n(r)\mu, \hj \pm \fjs, j}(z,\sigma)  = 
\mathcal{J}_{n(r)\mu \hj j}(z,\sigma) 
\end{gather}
\end{subequations}
from the natural periodicities of the eigenvalue problems.

%%MANU 2.8
The composite form and operator products of the eigenfields in terms of themselves are 
then straightforwardly computed from their definitions and the 
original operator products (2.1). We will not write them
out explicitly here (see however the remark after Eq.~(3.4)),
but call attention only to some useful quantities,
the \emph{twisted metrics}, which appear in the operator
products of the eigenfields:
%MANU 2.8'
\begin{subequations}
\label{2.8p.group1}
\begin{gather}
\label{2.8.1}
\begin{split}
\mathcal{G}_{\hj j; \hat{\ell} \ell}( \sigma ) &= 
\sqfjs \sqfls V( \sigma )\sS{\hj j}{I} V( \sigma )\sS{\hat{\ell} \ell}{J}\delta_{I J}
\\
& = \delta_{j\ell }\fjs \kd{\hj + \hat{\ell}}{0 \bmod \fjs},
\end{split}
\\
\label{2.8.2}
\begin{split}
\mathcal{G}_{n(r)\mu ; n(s)\nu} ( \sigma ) &= \chi_{n(r)\mu} ( \sigma ) \chi_{n(s)\nu}( \sigma )
U( \sigma )\sS{n(r)\mu}{a} U( \sigma )\sS{n(s)\nu}{b} G_{ab}\\
&= 
\kd{n(r) + n(s)}{0 \bmod \fjs}\mathcal{G}_{n(r)\mu ; -n(r)\nu}( \sigma )
\end{split}
\\
\label{2.8.3}
\sum_{n(t), \eta} \mathcal{G}^{n(r)\mu ; n(t)\eta} ( \sigma ) \mathcal{G}_{n(t)\eta ; n(s) \nu}( \sigma ) =
\delta\sS{n(r)\mu}{n(s)\nu}
\end{gather}
\end{subequations}
%MANU 2.8''
%In this paper, we do not choose any particular element $\omega( \sigma)$ 
%of any particular non-trivial $H'_{26}$ (see however Sec.~10). 
In what follows, the information about the choice of  $\omega( \sigma)  \in H'_{26}$ is encoded in the quantities $n(r)\mu$,
$\rho(\sigma)$ and the twisted metric $\mathcal{G}_{\cdot}(\sigma)$ and its 
inverse $\mathcal{G}^{\cdot}(\sigma)$ in Eqs.~(2.8b,c). The inverse of 
the metric (2.8a) is obtained by inverting the factor $f_{j}(\sigma)$, 
while the inverse of Eq.~(2.8b) involves the inverse of the normalizations
and replacement of the eigenmatrices by their adjoints.
%Since we are not choosing here any particular
%element of $\omega( \sigma)  \in H'_{26}$, the quantum numbers $n(r)\mu$ 
%and order $\rho(\sigma)$ of $\omega( \sigma) $,  
%as well as the twisted metric 
%$\mathcal{G}_{.}(\sigma)$ in Eq.~(2.8b) 
%and its inverse $\mathcal{G}^{.}(\sigma)$ in Eq.~(2.8c) will appear 
%frequently below.

%MANU 2.9
At this stage we have only rearranged 
the untwisted theory in terms of the eigenfields
$\mathcal{A}(z,\sigma) $. The final step in the orbifold program is
the transition to twisted sector~$\sigma$ of the orbifold by
an application of the \emph {principle of local isomorphisms} [9,11,12,15-17] %MANU 2.10
\begin{subequations}
\label{2.10.group1}
\begin{gather}
\label{2.10.1}
\mathcal{A} (z,\sigma)  \to \hat{A} (z,\sigma) ,
\\
\label{2.10.2}
\hbox{operator products of $\{\mathcal{A} (z,\sigma) \}$ } \to 
\hbox{ operator products of $\{\hat{A} (z,\sigma) \}$},
\\
\label{2.10.3}
\hbox{diagonal automorphic responses } \to
\hbox{ monodromies,}
\end{gather}
\end{subequations}
where $\{ \hat{A} (z,\sigma) \}$ are now the twisted fields of twisted
sector~$\sigma$ of the orbifold. The name of the principle derives
from part (b) of Eq.~(2.9), that the operator products of the twisted fields
are the same as (locally isomorphic to) the operator
products of the eigenfields. 
(We remind that there is another, equivalent way
around the commuting diagrams of Refs.~[9,11,17] to get from the untwisted 
fields to
the twisted fields. This path involves first a parallel application of
the principle of local isomorphisms, followed by a monodromy
decomposition to obtain the twisted fields $\hat{A}$.)

\section{The twisted operator products of sector~$\sigma$}\label{Section 3}
%%MANU 3.1
Having completed the steps above, we
emerge in the orbifold with the following twisted
stress tensors of sector~$\sigma$
\begin{equation}
\hat{\theta}_{\hj j} (z,\sigma) = \sum_{n(r)\mu\nu} \frac{\mathcal{G}^{n(r)\mu; -n(r)\nu}( \sigma )}{2\fjs}
%\hfill\cr \hfill{}\times % if needed for space
\sum_{\hat{\ell} = 0}^{\fjs-1} 
\no{\hat{J}_{n(r)\mu \hat{\ell} j} (z,\sigma)  \hat{J}_{-n(r),\nu,\hj-\hat{\ell}, j}(z,\sigma) }
\end{equation}
where $\no{\cdots}$ is now operator-product normal ordering in the orbifold.
The monodromies of these operators are
\begin{subequations}
\label{3.1.group1}
\begin{gather}
\label{3.1.1}
\hat{\theta}_{\hj j }(z \twone, \sigma ) = e^{-2\pi i \hatjovfs}\hat{\theta}_{\hj j}(z,\sigma) ,
\\
\label{3.1.2}
\hat{J}_{n(r)\mu \hj j}(z \twone, \sigma ) = e^{-2\pi i (\hatjovfs+\frac{n(r)}{\rho(\sigma) })}
\hat{J}_{n(r)\mu \hj j } (z,\sigma) 
\end{gather}
\end{subequations}
and the periodicities
\begin{subequations}
\label{3.1.group2}
\begin{gather}
\label{3.1.3}
\hat{\theta}_{\hj \pm \fjs, j}(z,\sigma) = \hat{\theta}_{\hj j}(z,\sigma),
\\
\label{3.1.4}
\hat{J}_{n(r)\pm\rho(\sigma),\mu \hj j}(z,\sigma) = \hat{J}_{n(r)\mu, \hj\pm \fjs,j}(z,\sigma)
= \hat{J}_{n(r)\mu \hj j}(z,\sigma)
\end{gather}
\end{subequations}
are inherited from the eigenfields.

%MANU 3.2
The operator products of sector~$\sigma$ are obtained as
\begin{subequations}
\label{3.2.group1}
\begin{gather}
\label{3.2.1}
\begin{gathered}[b]
\hat{\theta}_{\hj j}(z,\sigma) \hat{\theta}_{\hat{\ell} \ell}(\omega,\sigma) 
=
\delta_{j\ell}
\biggl[
\frac{\kd{\hj+\hat{\ell}}{0 \bmod \fjs}\frac{26}{2}\fjs}{(z - \omega)^4}\hskip8em{}\\
\hfill{}+
\Bigl(\mfrac{2}{(z-\omega^2)^2}+\mfrac{1}{z-\omega}\partial_\omega\Bigr)
\hat{\theta}_{\hj+\hat{\ell},j}(\omega,\sigma)\biggr]\quad\\
\hfill{}+
\no{\hat{\theta}_{\hj j}(z,\sigma) \hat{\theta}_{\hat{\ell} \ell}(\omega,\sigma) }
\end{gathered}
\\
\label{3.2.2}
\begin{gathered}[b]
\hat{\theta}_{\hj j}(z,\sigma) \hat{J}_{n(r)\mu\hat{\ell}\ell}(\omega,\sigma)
 = 
\delta_{j\ell}\Bigl(
\mfrac{1}{(z - \omega)^2}+ \mfrac{1}{z-\omega}\partial_{\omega}\Bigr)
\hat{J}_{n(r)\mu,\hj+\hat{\ell},j}(\omega,\sigma)\hskip3em{}\\
\hfill{}+
\no{\hat{\theta}_{\hj j}(z,\sigma) \hat{J}_{n(r)\mu\hat{\ell}\ell}(\omega,\sigma)}
\end{gathered}
\\
\label{3.2.3}
\begin{gathered}[b]
\hat{J}_{n(r)\mu \hj j}(z,\sigma) \hat{J}_{n(s)\nu\hat{\ell}\ell}(\omega,\sigma) 
= 
\delta_{j\ell}
\biggl(
\frac{\fjs \mathcal{G}_{n(r)\mu;n(s)\nu}( \sigma ) \kd{\hj +\hat{\ell}}{0 \bmod \fjs}}{(z-\omega)^2}\biggr)\hskip3em{}\\
\hfill{}+
\no{\hat{J}_{n(r)\mu \hj j}(z,\sigma) \hat{J}_{n(s)\nu\hat{\ell}\ell}(\omega,\sigma) }.
\end{gathered}
\end{gather}
\end{subequations}
The relations above provide a complete description of twisted sector~$\sigma$.
If desired, the previously-omitted details of the eigenfield system can 
be obtained from these statements by going backward $\hat{A} \to \mathcal{A}$, while
replacing the monodromies (3.2) with the diagonal
automorphic responses (2.6) of the eigenfields. We will
comment on applications to specific orbifold-string systems
after finding the corresponding mode algebras below.

\section{The twisted mode algebras of sector~$\sigma$}\label{Section 4}
%MANU 4.1
The twisted operator-product form of the system
above is straightforwardly translated to the mode-algebraic
description of the sector.  With attention to the monodromies (3.2)
and the  conformal-weight terms $(\Delta/(z-\omega)^2)$ in the operator products,
we define the modes of the stress tensors and twisted currents as follows:
\begin{subequations}
\label{4.1.group1}
\begin{gather}
\label{4.1.1}
\hat{\theta}_{\hj j} (z,\sigma) = \sum_{m\in\Z} \hat{L}_{\hj j}(m+\hatjovfs)z^{-(m+\hatjovfs)-2},
\\
\label{4.1.2}
\hat{J}_{n(r)\mu\hj j}(z,\sigma) = \sum_{m\in\Z}\hat{J}_{n(r)\mu\hj j} ( m + \nrovrhos +\hatjovfs)
z^{-(m+\nrovrhos +\hatjovfs) - 1}.
\end{gather}
\end{subequations}
%MANU 4.2
This gives immediately the mode form of 
the orbifold Virasoro generators
\begin{multline}
\hat{L}_{\hj j}(m+\hatjovfs) =
\sum_{n(r)\mu\nu}\frac{\mathcal{G}^{n(r)\mu; -n(r),\nu}( \sigma )}{2\fjs}
\sum_{\hat{\ell}=0}^{\fjs-1}\sum_{p\in\Z} \,\times \\
\times
\no{\hat{J}_{n(r)\mu\hat{\ell} j}(p+\nrovrhos +\hatlovfjs) 
\hat{J}_{-n(r),\nu,\hj-\hat{\ell},j}(m-p-\nrovrhos +\tfrac{\hj-\hat{\ell}}{\fjs})}
\end{multline}
and the mode periodicities
\begin{subequations}
\label{4.2.group1}
\begin{gather}
\label{4.2.1}
\hat{L}_{\hj \pm\fjs,j}(m+\tfrac{\hj\pm\fjs}{\fjs}) = \hat{L}_{\hj j}(m\pm 1 + \hatjovfs),\\
\label{4.2.2}
\begin{split}
\hat{J}_{n(r)\pm \rho(\sigma),\mu\hj j}(m +\tfrac{n(r)\pm\rho(\sigma)}{\rho(\sigma)} + \hatjovfs) & =
\hat{J}_{n(r)\mu,\hj\pm \fjs,j}(m+\nrovrhos +\tfrac{\hj\pm\fjs}{\fjs})\\
& =
\hat{J}_{n(r)\mu\hj j}(m\pm 1 + \nrovrhos +\hatjovfs).
\end{split}
\end{gather}
\end{subequations}
The operator-product normal-ordered forms (4.2) of the orbifold Virasoro generators are
not as useful as the mode-normal ordered forms we shall
obtain for these generators later.

%MANU 4.3
We give next the twisted mode algebras of sector $\sigma$
\begin{subequations}
\label{4.3.group1}
\begin{gather}
\label{4.3.1}
\begin{gathered}[b]
\bigl[ \hat{L}_{\hj j}(m+\hatjovfs),\hat{L}_{\hat{\ell}\ell}(n+\hatlovfs)\bigr]\hfill\\
\qquad{}
= \delta_{j\ell }\Bigl\{
(m-n+\tfrac{\hj-\hat{\ell}}{\fjs}) \hat{L}_{\hj-\hat{\ell},j}(m+n+\tfrac{\hj+\hat{\ell}}{\fjs})\hfill\\
\hskip8em{}
+\tfrac{26\fjs}{12}(m+\hatjovfs)\bigl((m+\hatjovfs)^2 - 1\bigr)
\kd{m+n+\frac{\hj+\hat{\ell}}{\fjs}}{0}\Bigr\},
\end{gathered}
\\
\label{4.3.2}
\begin{gathered}[b]
\bigl[
\hat{L}_{\hj j}(m+\hatjovfs), \hat{J}_{n(r)\mu \hat{\ell}\ell}(n+\nrovrhos +\hatlovfs)\bigr]\hfill\\
\hskip5em{}
= -\delta_{j\ell}(n+\nrovrhos  + \hatlovfjs) \hat{J}_{n(r)\mu, \hj+\hat{\ell}, j}
(m+n+\nrovrhos +\tfrac{\hj+\hat{\ell}}{\fjs}),
\end{gathered}
\\
\label{4.3.3}
\begin{gathered}[b]
\bigl[
\hat{J}_{n(r)\mu\hj j}(m+\nrovrhos + \hatjovfs), 
\hat{J}_{n(s)\nu\hat{\ell}\ell}(n+\tfrac{n(s)}{\rho(\sigma)} + \hatlovfs)\bigr]\hfill\\
\hskip3em{}
=\delta_{j\ell }\fjs (m +\nrovrhos + \hatjovfs) \kd{n(r)+n(s)}{0\bmod\rho(\sigma)}\hfill\\
\hskip13em{}
\times \kd{m+n+\frac{n(r)+n(s)}{\rho(\sigma)}+\frac{\hj+\hat{\ell}}{\fjs}}{0}
\mathcal{G}_{n(r)\mu;-n(r),\nu}( \sigma ),
\end{gathered}
\end{gather}
\end{subequations}
which are obtained by standard [11] orbifold contour methods
from the twisted operator products (3.4) and the mode expansions (4.1) of the operators. The reader will recognize
in particular the general orbifold Virasoro algebra (4.4a) obtained 
earlier${}^{\text{\footnotemark}}$  %TODO
\footnotetext{The general orbifold Virasoro algebra [4.4a] was first
obtained in the WZW permutation orbifolds [15] with $26\to c_g$, where
$c_g$ is the central charge of the affine-Sugawara construction [22] on Lie $g$.}
in Ref.~[6] and quoted in Eq.~(1.3) of the 
Introduction. We remind the reader of the ranges given in 
Eqs.~(2.3),(2.4) for the quantum numbers $\hat{j}j$ $(H({\rm perm})_{K})$ and 
$n(r)\mu$ $(H'_{26})$ which 
appear in this result, as well as the definition of the twisted metric $\mathcal{G}_{.}(\sigma)$
in Eq.~(2.8b).

%%MANU 4.4
%Let us turn next to a brief discussion 
%of the general orbifold Virasoro algebra \eqref{TODOeq} of sector~$\sigma$.
In further detail, the orbifold Virasoro algebra (4.4a) of sector $\sigma$ is semisimple with 
respect to the cycles $j$ of each $W( \sigma )\in H({\rm perm})_K)$, and each
cycle has its own integral Virasoro subalgebra at cycle central charge
$\hat{c}_j(\sigma) = 26\fjs$:
\begin{equation}
\label{4.4.1}
\begin{gathered}[b]{}
[ \hat{L}_{0j}(m),
\hat{L}_{0\ell}(n) ] = 
\delta_{j\ell}\bigl\{
(m-n) \hat{L}_{0j}(m+n) + \tfrac{26\fjs}{12} m(m^2-1)\delta_{m+n,0}\bigr\},\\
j,\ell = 0,1,\ldots,N( \sigma ) -1,
\end{gathered}
\end{equation}
where $N( \sigma )$ is the number of cycles in sector~$\sigma$.
% (these are
%the cycles in element $W( \sigma )\in H({\rm perm})_K)$. 
The total Virasoro generators of sector~$\sigma$ 
are obtained by summing over the cycles of the sector
\begin{subequations}
\label{4.4.group2}
\begin{align}
\label{4.4.2}
\hat{L}_{ \sigma }(m) &= {\tsty \sum\limits_j}\hat{L}_{0j}(m), 
\quad \hat{c} \ps = \tsty\sum\limits_j \hat{c}_j( \sigma ) = 26K 
\\
[\hat{L}_{ \sigma} (m) , \hat{L}_{\sigma}(n)] &= (m-n)
\hat{L}_{\sigma}(m+n) + \tfrac{26K}{12}m(m^2+1)\delta_{m+n,0},
\end{align}
\end{subequations}
where we have used the cycle sum rule in Eq.~(2.3c) to obtain the sector
central charges $\hat{c}(\sigma)$.

%MANU 4.5
Together, the twisted mode algebras (4.4) and the extended physical-state conditions 
(1.3a) form what we will call the \emph {general cycle dynamics} of the matter in  
cycle $j$ of sector $\sigma$. (The cycle dynamics includes 
the composite structure (4.2) of the orbifold Virasoro generators, but we 
remind that a more useful form of this structure will be obtained in the following section.)

We conclude with some comments on the
applicability of the general cycle dynamics to the sectors of the three families (1.1)
of orbifold-string theories of permutation-type. 
These results
are complete as they stand for the twisted open string systems, 
including  the open-string analogues (1.1b) 
of the generalized permutation orbifolds.  The open-string
sectors of the orientation orbifolds (1.1c) are included in the special
 case $K=f_0( \sigma ) = 2$ and $a_2 = 17/8$ of
these results at $\hat{c} ( \sigma ) = 52$. 
%The cycle dynamics of open-string sectors at  $\hat{c} ( \sigma ) = 52$
%were described earlier in Ref.~[3].
Right-mover copies of the cycle dynamics must be added 
to describe the generalized permutation orbifolds in Eq.~(1.1a).
The cycle dynamics of  both open- and closed-string  sectors at  $\hat{c} ( \sigma ) = 52$
were described earlier in Ref.~[3].
%For the generalized
%permutation orbifolds (1.1a), one needs to include
%a right-mover copy  -- including a right-mover copy of the extended physical
%state conditions -- of these results as well. 
Finally, the cycle dynamics of the
closed-string sectors of the orientation orbifolds (the ordinary 
space-time orbifolds $U(1)^{26}/H'_{26}$)
are obtained with a right-mover copy by choosing $K=f_0( \sigma )= 1$ and therefore 
the conventional intercept $a_1 = 1$ at $\hat{c} ( \sigma )=26$.

\section{Mode normal-ordering}\label{Section 5}
%MANU 5.1
%We are not yet finished with our description of the matter of
%the orbifold-string theories of permutation-type because the operator-normal-ordered
%form \eqref{TODOeq} of the orbifold Virasoro
%generators is not very practical for computations.

To obtain a more useful form of the orbifold Virasoro generators $\{L_{\hat{j}j}\}$, 
the next step in the orbifold program is the introduction
of \emph{mode normal-ordering}
%MANU 5.2
\begin{multline}
\label{5.2.1}
\nosub{
\hat{J}_{n(r)\mu \hj j}(m+\nrovrhos+\hatjovfs)
\hat{J}_{n(s)\nu\hat{\ell} \ell}(n + \tfrac{n(s)}{\rho(\sigma)} + \hatlovfs)
}{M}\\
{}\equiv
\phantom{{}+{}}
\theta((m+\nrovrhos+\hatjovfs)\geq 0)
\hat{J}_{n(s)\nu\hat{\ell}\ell}(n + \tfrac{n(s)}{\rho(\sigma)} + \hatlovfs)
\hat{J}_{n(r)\mu \hj j}(m+\nrovrhos+\hatjovfs)\\
{}
+
\theta((m+\nrovrhos+\hatjovfs) < 0)
\hat{J}_{n(r)\mu \hj j}(m+\nrovrhos+\hatjovfs)
\hat{J}_{n(s)\nu\hat{\ell} \ell}(n + \tfrac{n(s)}{\rho(\sigma)} + \hatlovfs)
\end{multline}
to replace the operator-product normal-ordering in Eq.~(4.2).

%MANU 5.3
The reordering is somewhat intricate,
so we will sketch the intermediate steps. From the 
definition (5.1) of mode normal-ordering and 
the commutator (4.4c) of two twisted currents, we obtain first
the relation between the product of two modes and the
mode normal-ordered product of the modes as follows:
%MANU 5.4
\begin{multline}
\label{5.4.1}
\hat{J}_{n(r)\mu \hj j}(m+\nrovrhos+\hatjovfs)
\hat{J}_{n(s)\nu\hat{\ell} \ell}(n + \nrovrhos + \hatlovfs)
\\
\hskip5em
{}
=
\nosub{
\hat{J}_{n(r)\mu \hj j}(m+\nrovrhos+\hatjovfs)
\hat{J}_{n(s)\nu\hat{\ell} \ell}(n + \nrovrhos + \hatlovfs)}{M}
\hfill
\\
\hskip5em
\phantom{{}={}}
{}+
\theta((m+\nrovrhos + \hatjovfs)\geq 0 )\delta_{j\ell}
\fjs (m+\nrovrhos + \hatjovfs)
\hfill
\\
{}\cdot
\kd{n(r)+n(s)}{0 \bmod\rho(\sigma)}
\kd{m+n+\frac{n(r)+n(s)}{\rho(\sigma)} + \frac{\hj+\hat{\ell}}{\fjs}}{0}
\mathcal{G}_{n(r)\mu;-n(r),\nu}( \sigma ).
\end{multline}
%MANU 5.5 
Then using the mode expansions and the $\hat{J}\hat{J}$ operator
product in Eq.~(3.4c), one straightforwardly obtains
the following exact relation between the two types
of normal-ordering of two local currents:
%MANU 5.6
\begin{subequations}
\begin{gather}
\begin{gathered}[b]
\no{
\hat{J}_{n(r)\mu\hj j}(z,\sigma) \hat{J}_{n(s)\nu\hat{\ell} \ell}(\omega,\sigma)}
-
\nosub{
\hat{J}_{n(r)\mu\hj j}(z,\sigma) \hat{J}_{n(s)\nu\hat{\ell} \ell}(\omega,\sigma)}{M}
\hskip8em
\hfill
\\
\qquad
{}= \fjs \delta_{j\ell} \mathcal{G}_{n(r)\mu;-n(r),\nu}( \sigma )
\kd{n(r)+n(s)}{0\bmod \rho(\sigma)}
\kd{\hj+\hat{\ell}}{0 \bmod \fjs}
\hfill
\\
\hskip4em
\cdot
\Bigl[
\mfrac{1}{z\omega}\Bigl(\mfrac{\omega}{z}\Bigr)^{ X_{\hj}}
\Bigl\{
X_{\hj}
\mfrac{z}{z-\omega}
(
\theta( 0 \leq X_{\hj} < 1) 
+ \mfrac{z}{w}
\theta( 1 \leq X_{\hj} < 2) 
)
\hfill
\\
\hfill
{}
+
\mfrac{z\omega}{(z-\omega)^2}
\theta( 0 \leq X_{\hj} < 1) 
- 
\mfrac{z(z-2\omega)}{\omega^2}
\theta( 1 \leq X_{\hj} < 2) \Bigl\}
\hskip1em
\\
\hfill{}
-\mfrac{1}{(z-\omega)^3}
\Bigr]
\end{gathered}
\\
X_{\hj}\equiv \mfrac{\bar{n} (r)}{\rho(\sigma)} + \mfrac{\bhj}{\fjs},
\quad
0\leq X_{\hj} < 2,
\quad
\forall
\bar{n} (r), \bhj.
\end{gather}
\end{subequations}
Here $\theta$ is the Heaviside function, and we have introduced the 
notational simplification $X_{\hj}\equiv X_{\hj n(r)}$. As $z\to\omega$, 
Eq.~(5.3) gives the exact relation between 
the two types of local normal-ordered current bilinears 
%MANU 5.7
\begin{multline}
\hskip2.5em
\no{
\hat{J}_{n(r)\mu\hj j}(z,\sigma) \hat{J}_{n(s)\nu\hat{\ell}\ell}(z,\sigma)}
=
\nosub{
\hat{J}_{n(r)\mu\hj j}(z,\sigma) \hat{J}_{n(s)\nu\hat{\ell}\ell}(z,\sigma)}{M}\\
\hskip-3em
{}
+\delta_{j\ell}\fjs 
\kd{\hj+\hat{\ell}}{0 \bmod \fjs}\mathcal{G}_{n(r)\mu;-n(r),\nu}( \sigma )
\kd{n(r)+n(s)}{0\bmod \rho(\sigma)}\\
{}
\cdot \mfrac{1}{z^2}\abs{ 1 - X_{\hj}}(1-\abs{1-X_{\hj}})
\hskip2.5em
\end{multline}
which is now in the desired form for application to the
orbifold Virasoro generators.

%MANU 5.8
With the relation (5.4), we can convert
the operator-product normal-ordered forms (3.1) or (4.2)
of the extended stress tensors and orbifold Virasoro generators to
the following mode-ordered forms
%MANU 5.9
\begin{subequations}
\label{5.9.group1}
\begin{gather}
\label{5.9.1}
\begin{gathered}[b]
\hat{\theta}_{\hj j}(z,\sigma) = \mfrac{1}{2\fjs}
\sum_{n(r)\mu\nu}
\mathcal{G}^{n(r)\mu;-n(r),\nu}( \sigma )
\sum_{\hat{\ell}=0}^{\fjs-1}
\nosub{\hat{J}_{n(r)\mu \hat{\ell} j}(z,\sigma) \hat{J}_{-n(r),\nu,\hj-\hat{\ell},j}(z,\sigma)}{M}\\
\hfill{}\!\!\!+
\frac{\kd{\hj}{0 \bmod \fjs}}{z^2}\hat{\Delta}_{0j}( \sigma )
\end{gathered}\\
\label{5.9.2}
\begin{gathered}[b]
\hat{L}_{\hj j}(m+\hatjovfs) = \mfrac{1}{2\fjs}
\sum_{n(r)\mu\nu}
\mathcal{G}^{n(r)\mu;-n(r),\nu}( \sigma )
\sum_{\hat{\ell}=0}^{\fjs-1}
\sum_{p\in\Z}\,\,\times\hskip12em\hfill\cr
\hfill{}
\times
\nosub{
\hat{J}_{n(r)\mu\hat{\ell} j}(p + \nrovrhos + \hatlovfjs)
\hat{J}_{-n(r)\nu,\hj-\hat{\ell},j}
(m-p-\nrovrhos + \tfrac{\hj-\hat{\ell}}{\fjs})}{M}
\\
\hfill{}\!\!\!+
\kd{m+\hatjovfs}{0}\hat{\Delta}_{0j}( \sigma )
\end{gathered}
\end{gather}
\end{subequations}
%MANU 5.10
where the shifts $\{\hat{\Delta}_{0j}\ps\}$, given below, will be called the
\emph{conformal weights of cycle~$j$ in sector~$\sigma$}. (These
conformal weights 
were called the partial conformal weights in the early examples of Ref.~[15].)

We give the explicit forms of the conformal weights $\{\hat{\Delta}_{0j}\ps\}$ in a
number of steps, which show various properties of
these quantities:
%MANU 5.11
\begin{subequations}
\label{5.11.group1}
\begin{align}
\label{5.11.1}
&
\begin{gathered}[b]
\hat{\Delta}_{0j}( \sigma ) = \mfrac 14 
\sum_{\substack{n(r)\mu}{n(s)\nu}}
\mathcal{G}^{n(r)\mu ; n(s)\nu} ( \sigma )
\mathcal{G}_{n(r)\mu ; n(s)\nu} ( \sigma )
\sum_{\hj=0}^{\fjs-1} \times\hfill\\
\hskip5em\quad{}\times
\{
\theta(0\leq X_{\hj}<1) 
X_{\hj} (1-X_{\hj})
+
\theta(1\leq X_{\hj}<2) 
(X_{\hj} - 1)(2 - X_{\hj})
\}
\end{gathered}
\\
\label{5.11.2}
&
= \mfrac14
\sum_{n(r)\mu}
\sum_{\hj = 0}^{\fjs - 1}
\{
\theta(0\leq X_{\hj}<1) 
X_{\hj}(1- X_{\hj})
+
\theta(1\leq X_{\hj}<2) 
(X_{\hj} - 1)(2 - X_{\hj})
\}
\\
\label{5.11.3}
&
\begin{gathered}[b]
=\mfrac12
\sum_r \dim [n(r)]
\sum_{\hj = 0}^{\fjs -1 }
(1 -\tfrac{\bar{n}(r)}{\rho(\sigma)} - \tfrac{\bhj}{\fjs})
%\hfill\\ \hskip10em\quad{}
\{
\tfrac12 (\tfrac{\bar{n}}{\rho(\sigma)}+\tfrac{\bhj}{\fjs}) -
\theta((\tfrac{\bar{n}}{\rho(\sigma)}+\tfrac{\bhj}{\fjs}) \geq 1)
\}.
\end{gathered}
\end{align}
\end{subequations}
%MANU 5.12
To obtain the second form (5.6b), we have used Eq.~(2.8c) to do the sum on
$n(s)\nu$, and the final step (5.6c) uses the degeneracy sum
$\sum_\mu = \dim [\bar{n}(r)]$ in Eq.~(2.4b).
The bars on the quantities $\hj$ can be ignored here because the fundamental
range is explicit in the summations. The second form in
particular shows that the conformal weight of cycle~$j$ is
nonnegative
\begin{equation}
\hat{\Delta}_{0j}( \sigma ) \geq 0
\end{equation}
and we shall find stronger lower bounds below.

%MANU 5.13
%The conformal weight of cycle~$j$ will enter our discussion at
%several points below, so
%we also include a number of other useful forms of this quantity
Other useful forms of the conformal weights of cycle $j$ include:
\begin{subequations}
\label{5.13.group1}
\begin{gather}
\label{5.13.1}
\hat{\Delta}_{0j} ( \sigma ) =
\mfrac{13}{12}\Bigl(\fjs - \mfrac{1}{\fjs}\Bigr)
+
\mfrac{1}{\fjs}\hat{\delta}_{0j}( \sigma ),\\
\label{5.13.2}
\tsty
\hat{\delta}_{0j}( \sigma ) \equiv
\frac{f_{j}(\sigma)}{2} \sum\limits_r \dim [\bar{n}(r)]
\hat{A} [\frac{\bar{n}(r)}{\rho(\sigma)}],
\\
\label{5.13.3}
\tsty
\sum\limits_r \dim [\bar{n}(r)] = 26,
\\
\label{5.13.4}
\begin{gathered}[b]
\tsty
\hat{A} [\frac{\bar{n}(r)}{\rho(\sigma)}] \equiv
\bigr(\frac{\bar{n}(r)}{\rho(\sigma)} - \frac{1}{\fjs}\bigl)(
\theta\bigl(\frac{\bar{n}(r)}{\rho(\sigma)} \geq \frac{1}{\fjs}\bigr) 
-
\frac{\fjs}{2}\frac{\bar{n}(r)}{\rho(\sigma)})\hskip9em\hfill\\
\hskip8em{}
+
\sum\limits_{\hj = 2}^{\fjs - 1} 
\tsty
(\frac{\bar{n}(r)}{\rho(\sigma)} - \frac{\hj}{\fjs})
\theta\bigl(\frac{\bar{n}(r)}{\rho(\sigma)} \geq \frac{\hj}{\fjs}\bigr) ,
\end{gathered}
\\
\label{5.13.5}
\hskip6em
{}=\tfrac{\fjs}{2}
\sum\limits_{\hj = 0}^{\fjs - 1} 
\tsty
(\frac{\bar{n}(r)}{\rho(\sigma)} - \frac{\hj}{\fjs})
(\frac{\hj+1}{\fjs} - \frac{\bar{n}(r)}{\rho(\sigma)})
\theta(\hatjovfs \leq \frac{\bar{n}(r)}{\rho(\sigma)} < \frac{\hj+1}{\fjs}).
\end{gather}
\end{subequations}
In what follows, we will refer to the quantity 
$\hat{\delta}_{0j}( \sigma )$ as
%MANU 5.14
the \emph{conformal-weight shift} of cycle~$j$ in sector~$\sigma$.
The first form of the function $\hat{A}$ in Eq.~(5.8d) follows directly from Eq.~(5.6c), and is 
easier to evaluate explicitly for small cycle length $\fjs$. The second form of
$\hat{A}$ in Eq.~(5.8e) follows by induction from the first form, and shows that
\begin{equation}
\label{5.14.1}
\hat{A}[\tfrac{\bar{n}(r)}{\rho(\sigma)}] \geq 0
\quad\implies\quad
\hat{\delta}_{0j} ( \sigma ) \geq 0
\quad\implies\quad
\hat{\Delta}_{0j} ( \sigma ) \geq \mfrac{13}{12}\Bigl(\fjs -\mfrac{1}{\fjs}\Bigr).
\end{equation}
Moreover $\hat{A}$ is a continuous function, periodic in the 
$H'_{26}$-fraction $(\unfrac{\bar{n}}{\rho})$
with period $(\unfrac{1}{\fjs})$. The only zeroes of $\hat{A}$ are at the values
\begin{equation}
\label{5.14.2}
\fjs \tfrac{\bar{n}(r)}{\rho(\sigma)}\in \Z_{\geq 0}
\end{equation}
and the maximum value in each cell is $(1/\;8\fjs)$.

%MANU 5.15
Let us check our general result (5.8) for
the previously-studied cases [3] of $H({\rm perm})_2=\Z_2$ or $\Z_s({\rm w.s.})$,
i.e. for the generalized $\Z_2$-permutation orbifolds or the
twisted open-string sectors of the orientation orbifolds.
Choosing for either case the single nontrivial element
with a single cycle $j=0$ of length $f_0( \sigma ) = 2$,
we can easily$^{\text{\footnotemark}}$\footnotetext{On the other hand, we find a repeated
typo in Eqs.~(3.36c) and (3.38b) of the earlier Ref.~[18]:
The terms $(n(r)/2\rho(\sigma))^2$ in each of these equations should read simply
$(n(r)/\rho(\sigma))^2$, without the 2 in the denominator.}
do the sum explicitly over $\bhj=0,1$ to obtain
\begin{equation}
\label{5.15.1}
\tsty
\hat{\Delta}_{00}( \sigma ) = \frac{13}{8} +
\sum\limits_r \dim [\bar{n}(r)] (\frac{\bar{n}(r)}{\rho(\sigma)} - \frac{1}{2})
((\theta (\frac{\bar{n}(r)}{\rho(\sigma)} \geq \frac12)- \frac{\bar{n}(r)}{\rho(\sigma)})\geq\frac{13}{8}).
\end{equation}
This is in agreement with the result in Eq.~(2.3d) of Ref.~[3].

%MANU 5.16 
There is one more relation between the 
quantities discussed here which will be useful to record
\begin{equation}
\hat{\Delta}_{0j}( \sigma ) - \hat{a} _{\fjs} = \tfrac{1}{\fjs} (\hat{\delta} _{0j} ( \sigma ) - 1)
\end{equation}
where $\hat{a}_{\fjs}$ is the \emph{intercept} of cycle~$j$ in sector~$\sigma$ in
the extended physical-state condition (1.3a).

%MANU 5.17
We close this section with a simple
application of our results here to the \emph{twist-field state}
$\ket{0}_{j \sigma }$ of cycle~$j$ in sector~$\sigma$ 
\begin{subequations}
\label{5.17.group1}
\begin{gather}
\label{5.17.1}
\hat{J}_{n(r)\mu \hj j} ((m+\nrovrhos + \hatjovfs)\geq 0)
\ket{0}_{j \sigma } = 0,\\
\label{5.17.2}
\{\hat{L}_{\hj j}((m+\hatjovfs)\geq 0) - \hat{\Delta}_{0j}( \sigma )
\kd{m+\hatjovfs}{0}\} 
\ket{0}_{j \sigma } = 0,\\
\label{5.17.3}
(\hat{L} _{ \sigma }(m\geq 0) - \hat{\Delta}_{ \sigma }\kd{m}{0})
\ket{0}_{\sigma} = 0,\\
\label{5.17.4}
\tsty
\ket{0}_{ \sigma } \equiv \bigotimes_{j} \ket{0}_{j \sigma },
\quad
\hat{\Delta}_{ \sigma } \equiv \sum_j \hat{\Delta}_{0j}( \sigma ).
\end{gather}
\end{subequations}
The first line in Eq.~(5.13) defines this state, while the succeeding lines
then follow from the mode-normal-ordered form (5.5b) of the orbifold
Virasoro generators. Although the twist-field state is closely related to
the physical ground-state of each cycle (see Sec.~7), we should emphasize 
that the twist-field  state itself is not generally a physical state.
%which follows from the mode-ordered form of
%$\{\hat{L} _{\hj j} \}$ in Eq.~\eqref{TODOeq}.

\section{First discussion of the zero modes}\label{Section 6}
%MANU 6.1
To analyze the spectral problems associated
to the extended physical-state conditions (1.3a), we need
to separate out the \emph{zero modes} $\{\hat{J}_j(0)_ \sigma \}$ of
cycle~$j$ in sector~$\sigma$ from the doubly-twisted currents
\[
\hat{J}_{n(r)\mu\hj j}(m+\nrovrhos+\hatjovfs).
\]
We remind that $\mu = \mu(n(r))$ is the degeneracy index of the spectral
index $n(r)$ of each element of the 26-dimensional automorphism group $H'_{26}$, while $\{\hat{j}j\}$  
record the cycle-basis of each element of $H({\rm perm})_{K}$.
The zero modes are special cases of what we will call the \emph{integer-moded
sequences} $\{\hat{J}_j(m)_ \sigma \}$ of cycle~$j$ in sector~$\sigma$.

From the twisted current algebra (4.4c), we know that the zero modes
commute with all the currents, including themselves
\begin{equation}
\label{6.1.1}
[ \{ \hat{J}_j(0)_ \sigma \}, \hat{J}_{n(r)\mu \hat{\ell} \ell}(m+\nrovrhos+\hatlovfs)] = 0,
\quad\forall j,\ell, \sigma
\end{equation}
so we may alternatively refer to the zero modes as the
\emph{momenta} of cycle~$j$ in sector~$\sigma$. Then it is  natural to
define the number of zero modes in each cycle as
the \emph{target space-time dimension} of cycle~$j$ in sector~$\sigma$:
\begin{equation}
\label{6.1.2}
\hat{D}_j \ps \equiv \dim \{\hat{J}_j(0)_ \sigma \}.
\end{equation}
%MANU 6.2
The target space-time interpretation 
is a topic of central importance in the new string theories. 
In this paper, we confine ourselves
only to general properties of the space-time
dimensions, across all the bosonic prototypes (1.1) of the orbifold-string theories
of permutation-type. The general formulae developed 
here are however quite model-dependent,
involving the choice of subgroup $H'_{26}$ and the particular element $\omega \ps\in H'_{26}$.
Beyond the simple examples of trivial $H'_{26}$ in Sec.~10, we will 
return to study the target space-times of large classes
of specific models in succeeding papers of this series.

%MANU 6.3
From the total mode number of the doubly-twisted currents, it is straightforward 
to determine that the following conditions are necessary and sufficient
for integer-moded sequences and hence zero 
modes in cycle~$j$ of sector~$\sigma$: 
\begin{subequations}
\label{6.3.group1}
\begin{gather}
\label{6.3.1}
\{ \hat{J}_j (0)_ \sigma \} : \quad \fjs\tfrac{\bar{n}(r)}{\rho(\sigma)}
\in \Z_{\geq 0},\\
\label{6.3.2}
\bar{n}(r) = \tfrac{\rho(\sigma) \hj'}{\fjs}
\quad\hbox{such that }
{\hj}' = 0,1,\dots,\fjs-1, \ 
\bar{n}(r)\in \{0,1,\dots,\rho(\sigma)-1\}.
\end{gather}
\end{subequations}
The conditions in \eqref{6.3.2} are a more detailed
statement of the condition in \eqref{6.3.1}. The explicit
solutions of these conditions depend on
the cycle length $\fjs$ in sector $\sigma$ of $H({\rm perm})_K$, 
as well as the details of $\omega \ps\in H'_{26}$
reflected in the ratio $\frac{\bar{n}(r)}{\rho(\sigma)}$.
We emphasize that the condition (6.3a)
is the same condition under which the function 
$\hat{A} [\frac{\bar{n}(r)}{\rho(\sigma)}]=0$
(see Eq.~(5.10)), so that the integer-moded sequences
do not contribute to the conformal-weight shift
$\hat{\delta}_{0j}\ps$ of cycle~$j$ in sector~$\sigma$. This phenomenon
is familiar in ordinary untwisted string theory, where all sequences are integer-moded
and there are no conformal-weight shifts.

%MANU 6.4
Using the conditions (6.3), we can give a qualitative sketch of the integer-moded
sequences and zero modes as follows. We begin by noting that there are exactly two
possible types of such sequences, the first of which %
\begin{equation}%TODO
\hbox{type I:}\quad
\hat{J}_{0\mu(0)0j} (m) \xrightarrow[m=0]{}\hat{J}_{0\mu(0)0j}(0)
\end{equation}
is found if and only if $\bar{n} = 0$ occurs in the spectrum
of $\omega\ps\in H'_{26}$. This type occurs for example (see Sec.~10) in each cycle of 
every sector of the ``pure'' permutation orbifolds with trivial $H'_{26}$. 
The second, distinct type that can arise is the following family:
\begin{subequations}
\begin{gather}
\hbox{type II:}\quad
\hat{J}_{\frac{\rho(\sigma) \hj'}{\fjs}, \mu(\frac{\rho(\sigma)\hj'}{\fjs}),\fjs-\hj',j}(m+1)
\xrightarrow[m=-1]{}
\hat{J}_{\frac{\rho(\sigma) \hj'}{\fjs},\mu(\frac{\rho(\sigma) \hj'}{\fjs}),\fjs-\hj',j}(0),\hfill\\
\hj'=1,\dots,\fjs-1.
\end{gather}
\end{subequations}
Because the spectral index $\bar{n}(r)$ is an integer, this family  occurs only
when $\rho(\sigma)$ is a multiple of $\fjs$ or vice-versa. Examples
of this type with $(\fjs = 2, \rho(\sigma) = {\rm even})$ have been discussed
in Ref.~[3], and we will discuss both of these types more
systematically in succeeding papers.

%MANU 6.5
For our discussion below, it will be convenient to
introduce a formal Heaviside function for the 
momenta:
\begin{equation}
\label{6.5.1}
\theta\{ \hat{J}_j(0)_{ \sigma }\} \equiv
\begin{tcases}
1&\hbox{when $\omega( \sigma) \in H'_{26}$ allows the zero mode in $j \sigma $},\\
0&\hbox{otherwise}.
\end{tcases}
\end{equation}
This allows us to write formal expressions for the number
of target space-time dimensions and the
``momentum-squared'' operator of cycle~$j$ in sector~$\sigma$:
\begin{subequations}
\label{6.5.group1}
\begin{gather}
\label{6.5.2}
\hat{D}_j \ps =
\dim \{ \hat{J}_j (0)_{\sigma} = 
\sum_{\bar{n}(r)\mu\hj'}\theta\{\hat{J}_j(0)_{\sigma}\}\\
\label{6.5.3}
\begin{gathered}[b]
\hat{P}_j^2 \ps \equiv 
-\sum_{\mu,\nu}\theta \{\hat{J}_j (0)_{\sigma}\}
\bigl\{ \mathcal{G}^{0\mu ;0\nu} \ps \hat{J}_{0\mu0j} (0) \hat{J}_{0\nu0j} (0)\hfill\\
\hskip5em{}
+\sum_{\hj'=1}^{\fjs-1}\mathcal{G}^{\frac{\rho \hj'}{\fj},\mu;-\frac{\rho\hj'}{\fj},\nu}\ps
\hat{J}_{\frac{\rho \hj'}{\fj},\mu,\fj-\hj',j}(0)
\hat{J}_{-\frac{\rho\hj'}{\fj},\nu,\hj'-\fj,j}(0)\bigr\}.
\end{gathered}
\end{gather}
\end{subequations}
For brevity, we have omitted here the sector label $\sigma$ in both 
$f_{j}(\sigma)$ and $\rho(\sigma)$. The momentum-squared operator in Eq.~(6.7b) will play a central role
in the following analysis of the extended physical-state
conditions.

\section{The extended physical-state conditions}\label{Section 7}
%MANU 7.1
We begin this section by recalling the extended physical-state conditions for
the open-string analogues of the generalized permutation orbifolds
\begin{equation}
\label{7.1.1}
\Bigl[\frac{{\rm U}(1)^{26K}}{H_+}\Bigr]_{\rm open},\quad
H_+\subset H({\rm perm})_K \times H'_{26},
\end{equation}
which include the twisted open-string sectors of the orientation
orbifolds when $K=2$. As obtained in the BRST quantization of Ref.~[6] and quoted
above in Eq.~(1.3), these conditions read
\begin{subequations}
\label{7.1.group1}
\begin{gather}
\label{7.1.2}
(\hat{L}_{\hj j} ((m+\hatjovfs)\geq 0) - \hat{a}_{\fjs} \kd{m+\hatjovfs}{0})
\ket {\chi\ps}_j = 0,\\
\label{7.1.3}
\hat{c}_j\ps = 26\fjs,\quad
\hat{a}_{\fjs} = \frac{13 f_j^2\ps - 1}{12 \fjs}\\
\label{7.1.4}
\bhj = 0,1,\dots,\fjs-1,\quad 
j=0,1,\dots,N\ps -1
\end{gather}
\end{subequations}
where $\hat{c}_j\ps$ and $\hat{a}_{\fjs}$ are respectively the cycle 
central charge and the intercept of cycle $j$ in sector $\sigma$.
The integer $N(\sigma)$ in Eq.~(7.2c) is the number of cycles in sector $\sigma$,
while the orbifold Virasoro generators $\{\hat{L}_{\hj j}\}$ should now 
be taken in the mode normal-ordered form (5.5b).
%and $N\ps$ is the number of cycles in sector~$\sigma$. 

For each cycle~$j$ in every sector~$\sigma$, this system can be decomposed
into the extended gauge conditions
\begin{equation}
\label{7.1.5}
\hat{L}_{\hj j}((m + \hatjovfs)>0)
\kets{\chi\ps}{j} = 0
\end{equation}
%MANU 7.2
and the spectral subproblem for cycle~$j$ of sector~$\sigma$: 
\begin{subequations}
\label{7.2.group1}
\begin{gather}
\label{7.2.1}
(\hat{L}_{0j}(0) - \hat{a}_{\fjs})
\kets{\chi\ps}{j} = 0,\\
\label{7.2.2}
\hat{L}_{0j}(0) = \mfrac{1}{2\fjs}
( -\hat{P}^2_j\ps+\hat{R}_j\ps) +
\hat{\Delta}_{0j}\ps,\\
\label{7.2.3}
\begin{gathered}[b]
\hat{R}_j \ps \equiv
\sideset{}{'}\sum_{n(r)\mu\nu}
\mathcal{G}^{n(r)\mu;-n(r)\nu}\ps \sum_{\hj=0}^{\fjs-1}\sum_{p\in\Z} \times
\hfill
\\
\hskip5em{}
\times
\nosub{
\hat{J}_{n(r)\mu\hj j} (p +\nrovrhos + \hatjovfs)
\hat{J}_{-n(r),\nu,-\hj,j} (-p -\nrovrhos -\hatjovfs)
}{M}.
\end{gathered}
\end{gather}
\end{subequations}
The momentum-squared operator (6.7b) appears now among the terms of Eq.~(7.4b),
and the primed sum in the generalized
number operator $\hat{R}_j\ps$ denotes omission
of the zero modes.
%of $\hat{L}_{0j} (0)$.

%MANU 7.3
So long as the cycle-momenta $\{\hat{J}_j(0)_\sigma\}$
are not an empty set, the \emph {physical ground-state} of cycle~$j$
in sector~$\sigma$ is the $\{\hat{J}_j(0)_\sigma\}$-boosted twist-field
state $\ket{0,\hat{J}_j(0)}_\sigma$
\begin{subequations}
\label{7.3.group1}
\begin{gather}
\label{7.3.1}
\hat{J}_{n(r)\mu\hj j} ((m+\nrovrhos + \hatjovfs) > 0)
\ket{0,\hat{J}_j(0)}_\sigma 
= 0,
\\
\label{7.3.2}
\hat{R}_j\ps 
\ket{0,\hat{J}_j(0)}_\sigma 
= 
\hat{L}_{\hj j}((m+\hatjovfs)>0)
\ket{0,\hat{J}_j(0)}_\sigma 
= 0,\\
\label{7.3.3}
\hat{P}^2_j\ps
\ket{0,\hat{J}_j(0)}_\sigma 
=
\hat{P}^2_j\ps_{(0)}
\ket{0,\hat{J}_j(0)}_\sigma, \\
\label{7.3.4}
\hat{P}^2_j\ps_{(0)}
= 2 (\hat{\delta}_{0j}\ps -1)\geq -2
\end{gather}
\end{subequations}
where the \emph{ground-state momentum-squared} of
cycle~$j$ in sector~$\sigma$ is given in Eq.~\eqref{7.3.4}. To 
obtain this result, we 
used Eqs.~(7.4a,b), and the relation (5.12) between
the fundamental constants of the cycle. The explicit form of the conformal-weight
shift $\hat{\delta}_{0j}\ps\geq 0$ is given in Eq.~(5.8b), and we
see from Eq.~(7.5d) that the conformal-weight shift does indeed measure a 
shift from the ground-state momentum-squared $P^2 = -2$ of
an ordinary untwisted open string.

%MANU 7.4
According to Eqs.~\eqref{4.3.2} and \eqref{7.2.2}, the excited
states of cycle~$j$ in sector~$\sigma$ 
exhibit the \emph{level-spacing}
\begin{equation}
\label{7.4.1}
\Delta (\hat{P}^2_j\ps) = \Delta (\hat{R}_j\ps)
= 2\fjs \babs{ m +\nrovrhos +\hatjovfs}.
\end{equation}
These are the increments of mass-squared associated to the addition of any negatively-moded
current
\[
\hat{J}_{n(r)\mu\hj j} ((m+\nrovrhos+\hatjovfs)< 0)
\]
to products of other such currents on the ground-state
of cycle~$j$ in sector~$\sigma$.

%MANU 7.5
We turn next to the extended physical-state
conditions of all the closed-string sectors of the
generalized permutation orbifolds
\begin{equation}
\frac{{\rm U}(1)^{26K}}{H_+}
,\quad
H_+ \subset H({\rm perm})_K \times H'_{26}
\end{equation}
whose cycles also live at cycle central charge
$\hat{c}_j\ps = 26\fjs$ and sector central charge
$\hat{c}\ps = 26K$.
In these cases we have a left- and right-mover
copy of the extended physical-state conditions (7.2),
which we write as 
\begin{subequations}
\label{7.5.group1}
\begin{gather}
\label{7.5.1}
\begin{gathered}[b]
\hat{L}^L_{\hj j}((m+\hatjovfs)\geq 0)
\kets{\chi\ps}{j} = 
\hat{L}^R_{\hj j}((m+\hatjovfs)\geq 0)
\kets{\chi\ps}{j}
\\
 = 
\hat{a}_{\fjs} \kd{m+\hatjovfs}{0}
\kets{\chi\ps}{j},
\end{gathered}
\\
\label{7.5.2}
\hat{a}_{\fjs} = \frac{13 f_j^2\ps-1}{12\fjs},\hskip9em
\\
\label{7.5.3}
\hat{L}^L_{0j}(0) = \mfrac{1}{2\fjs}
(\hat{P}^2_{j}\ps^L +\hat{R}_j\ps^L +\hat{\Delta}_{0j}\ps),
\\
\label{7.5.4}
\hat{L}^R_{0j}(0) = \mfrac{1}{2\fjs}
(\hat{P}^2_{j}\ps^R +\hat{R}_j\ps^R +\hat{\Delta}_{0j}\ps).
\end{gather}
\end{subequations}
The extended Virasoro generators
$\{\hat{L}^L\}$ and $\{\hat{L}^R\}$ involve
the twisted left- and right-mover currents
$\{\hat{J}^L\}$ and $\{\hat{J}^R\}$ respectively.

%MANU 7.6
Following Ref.~[3], we study only the case
of decompactified zero modes, with the ordinary left-right
identifications:
\begin{subequations}
\label{7.6.group1}
\begin{gather}
\label{7.6.1}
\hat{J}_j^R (0)_\sigma 
= 
\hat{J}_j^L (0)_\sigma
=
\tfrac{1}{\sqrt{2}}\hat{J}_j(0)_\sigma,
\\
\label{7.6.2}
\hat{P}_j^2 \ps^R
= 
\hat{P}_j^2 \ps^L
=
\tfrac12 \hat{P}^2_j \ps.
\end{gather}
\end{subequations}
The closed-string momenta $\{\hat{J}_j(0)_\sigma\}$ and momentum-squared $\hat{P}^2_j \ps$ appear 
on the right side of Eqs.~(7.9 a,b), and $\hat{P}^2_j \ps$ has exactly the form (6.7b)
when expressed in terms  of $\{\hat{J}_j(0)_\sigma\}$.
%in \eqref{7.6.2} is defined to be in exactly the same
%form as given in Eq.~(6.7b).
Then the extended physical-state conditions (7.8) can be put in the form
%of cycle~$j$
%of each twisted closed string $ \sigma $ can be put in the form
\begin{subequations}
\label{7.6.group2}
\begin{gather}
\label{7.6.3}
\hat{P}_j^2 \ps
\kets{\chi\ps}{j} = 
2(\hat{P}_j^2\ps_{(0)} + \hat{R}_j^L\ps)
\kets{\chi\ps}{j}, \\
\label{7.6.4}
(\hat{R}^R_j\ps - \hat{R}^L_j\ps )
\kets{\chi\ps}{j} = 0,\\
\label{7.6.5}
\hat{L}^R_{\hj j} ((m+\hatjovfs)>0)
\kets{\chi\ps}{j} = 
\hat{L}^L_{\hj j} ((m+\hatjovfs)>0)
\kets{\chi\ps}{j} = 0
\end{gather}
\end{subequations}
where $\hat{P}^2_j\ps_{(0)}$ is defined in Eq.~\eqref{7.3.4}, and 
Eq.~(7.10b) is the level-matching condition for
cycle~$j$ in sector~$\sigma$.  Assuming again
that the momenta are not an empty set, we find
that the physical closed-string ground-state 
$\ket{0,\hat{J}_j(0)}_{\sigma}$
of cycle~$j$ has ground-state momentum-squared
%MANU 7.7
\begin{equation}
\label{7.7.1}
\hat{P}^2_j\ps^{\text{closed}}{(0)} = 
2 \hat{P}^2_j\ps_{(0)} = 4(\hat{\delta}_{0j}\ps - 1) \geq -4
\end{equation}
that is, twice the ground-state momentum-squared of the corresponding
twisted open string.

In what follows we will explicitly discuss
only the twisted open-string cases, but the corresponding closed-string
cases can easily be obtained from the results above. 
%including
%twice the ground-state momentum squared as shown in Eq.~\eqref{7.7.1}.

\section{The reduced formulation at $\cjs=26$}\label{Section 8}
%MANU 8.1
%Recall that the $\hat{c} = 52$ spectra of Refs.~\tempcite{} had
%an equivalent, reduced formulation at $c=26$. Here
%we shall find that such equivalent descriptions exists also for all cycles $j$ in
%all sectors $\sigma$ of the general orbifold-string theory
%of permutation-type, now at reduced central charge
%$c_j\ps = 26$ for each cycle~$j$ in sector~$\sigma$. As
%earlier, we emphasize that this equivalence holds only for
Recall that the $\hat{c}(\sigma) = 52$ physical spectral problems have an equivalent,
reduced description [3,4] at reduced central charge $c(\sigma)=26$. In 
this and the following section, 
we generalize this result to include all the bosonic prototypes (1.1) at 
sector central charge $\hat{c}(\sigma)=26K$ and cycle central charge $\hat{c}_{j}(\sigma)=26f_{j}(\sigma)$. In particular, we 
find the equivalent, reduced formulation of each cycle $j$ at \emph {reduced cycle central charge}  $c_j\ps = 26$, independent of the cycle. 
We emphasize with the earlier references that \emph{this equivalence holds only for
the cycle dynamics of the orbifold-string theories as described by the extended
physical-state conditions}, and \emph{not} for the underlying
orbifold CFT's themselves. One advantage of the original description at $\hat{c}(\sigma)=26K$
is locality [1], which provides twisted local vertex operators [13,15,18,21,4,5]. With 
Ref.~[3] we shall see however that the zero modes and target 
space-time dimensions are \emph {invariant} under the reduction, and
we shall emphasize in succeeding papers that the the target space-time 
properties of the theories are more easily studied in the reduced description.
%is particularly valuable in 
%target space-time dimensions, are \emph{invariant} under the reduction,
%We shall see that in either
%description, the zero mode structure, i.e.\ the 
%target space-time dimensions, are \emph{invariant} under the reduction,
%and moreover, as we shall emphasize in the following two papers,
%the reduced description is in many ways more transparent, especially
%with regard to the target space-time interpretation of the new string theories.

%MANU 8.2
In this paper we organize the discussion of the equivalent, reduced formulation
into two parts. In the present section we work out the operators of the reduced formulation
at $c_j\ps = 26$, leaving for Sec.~9 the reduced physical-state 
conditions and the equivalence of the two formulations at the string level.

The reduced (unhatted) operators of cycle~$j$ in sector~$\sigma$ are 
defined by the following map
\begin{subequations}
\label{8.2.group1}
\begin{gather}
\label{8.2.1}
L_j(M_j) \equiv 
\fjs \hat{L}_{\hj j}(m+\hatjovfs) - \tfrac{13}{12}(\fjs^2 -1)\kd{m+\hatjovfs}{0},
\\
\label{8.2.2}
J_{n(r)\mu j}(M_j + \fjs\tfrac{n(r)}{\rho(\sigma)})\equiv
\hat{J}_{n(r)\mu \hj j}(m+\nrovrhos +\hatjovfs),
\\
\label{8.2.3}
M_j \equiv \fjs m + \bhj\in\Z
\end{gather}
\end{subequations}
in terms of the hatted operators above. Since $m\in\Z$ and
$\bhj\in 0,1,\dots,\fjs-1$, the capitalized quantities $M_j$ cover the integers
once for each cycle~$j$, and indeed the map is one-to-one
at each fixed $(j, \sigma )$.
This map is in fact a modest generalization of the 
(inverse of) the order-$\lambda$ orbifold-induction procedure
of Borisov, Halpern, and Schweigert [7].

%MANU 8.3 
Then we find from Eq.~(4.4) the explicit algebra 
of the reduced operators:${}^{\text{\footnotemark}}$\footnotetext{
In Eq.~(4.2e) of Ref.~[3],
there is a missing factor $(M+2\nrovrhos)$,
which is now included in \eqref{8.3.3} when $\fjs = 2$.}
\begin{subequations}
\label{8.3.group1}
\begin{gather}
\label{8.3.1}
[L_j(M),L_\ell(N)] = \delta_{j\ell}
\{ (M-N)L_j(M+N) + \tfrac{26}{12} M(M^2 -1)\kd{M+N}{0}\},
\\
\label{8.3.2}
[L_j(M),
J_{n(r)\mu \ell}(N+ f_\ell\ps\nrovrhos) ]
=
-\delta_{j\ell}(N+\fls \nrovrhos)J_{n(r)\mu \ell}(M+N+\fls\tfrac{n(r)}{\rho(\sigma)}),
\\
\label{8.3.3}
\begin{gathered}[b]
{}
[
J_{n(r)\mu j}(M+ \fjs\nrovrhos),
J_{n(s)\nu \ell}(N+ f_\ell\ps\nsovrhos)]\hfill\\
\hskip 1em{}=
\delta_{j\ell}(M+\fjs \nrovrhos)\kd{n(r)+n(s)}{0\bmod\rho(\sigma)}
\kd{M+N+\fjs\smash[t]{\tfrac{n(r)+n(s)}{\rho(\sigma)}}}{0}
\,
\mathcal{G}_{n(r)\mu;-n(r)\nu}\ps,
\end{gathered}
\\
\label{8.3.4}
J_{n(r)\pm \rho(\sigma),\mu j}(m+\fjs\tfrac{n(r)\pm\rho(\sigma)}{\rho(\sigma)})=
J_{n(r)\mu}(m\pm\fjs+\fjs\tfrac{n(r)}{\rho(\sigma)}).
\end{gather}
\end{subequations}
In particular, Eq.~(8.2a) shows that the reduced generators $\{L_{j}(M)\}$ 
satisfy an ordinary Virasoro algebra with reduced cycle central charge $c_j\ps=26$ for
each cycle~$j$ in  any sector~$\sigma$. The total reduced
Virasoro generators of sector~$\sigma$
\begin{equation}
\label{8.3.5}
L_\sigma (M) \equiv \sum_j L_j(M)
\end{equation}
are then also Virasoro with reduced sector central charge
\begin{equation}
\label{8.3.6}
c(\sigma) = \sum_j c_j\ps = 26 N \ps
\end{equation}
where  $N\ps$ is the number of cycles in sector~$\sigma$.

%MANU 8.4
With Eq.~(5.5b), the map also gives the explicit form
of the reduced Virasoro generators of each cycle at
$c_j\ps=26$ in terms of the reduced currents:
\begin{subequations}
\label{8.4.group1}
\begin{gather}
\begin{gathered}[b]
\label{8.4.1}
L_j(M) = \delta_{M,0}\hat{\delta}_{0j}\ps
+ \tfrac12\sum_{n(r)\mu\nu} \mathcal{G}^{n(r)\mu;-n(r)\nu}\ps
\sum_{p\in\Z}\times\hfill\\
\hskip5em{}\times
\nosub{
J_{n(r)\mu j}(P + \fjs\nrovrhos)
J_{-n(r),\nu j}
(M-P-\fjs\nrovrhos)}{M},
\end{gathered}
\\
\label{8.4.2}
\begin{gathered}[b]
\hat{\delta}_{0j}\ps =
\tfrac12 \sum_r \dim [\bar{n}(r)]
\Bigl\{
(\fjs\tfrac{\bar{n}(r)}{\rho(\sigma)} - 1)(
\theta(\fjs\tfrac{\bar{n}(r)}{\rho(\sigma)}\geq 1) 
- \fjs\tfrac{\bar{n}(r)}{\rho(\sigma)})\hskip5.5em\\
\hskip15em{}
+\sum_{\hj = 2}^{\fjs -1}
(\fjs\tfrac{\bar{n}(r)}{\rho(\sigma)} -\hj)
\theta(\fjs\tfrac{\bar{n}(r)}{\rho(\sigma)}\geq \hj) \Bigr\},
\end{gathered}
\\
\label{8.4.3}
\hskip1em{}
=
\tfrac14 \sum_r\dim [\bar{n}(r)]
\sum_{\hj = 0}^{\fjs - 1}
(\fjs\tfrac{\bar{n}(r)}{\rho(\sigma)} -\hj)
(\hj + 1 - \tfrac{\fjs \bar{n}(r)}{\rho(\sigma)})
\theta(\hj\leq \tfrac{\fjs \bar{n}(r)}{\rho(\sigma)} < \hj + 1),\\
j=0,1,\dots,N\ps -1,\quad
\tsty\sum_j\fjs = K,\quad
\sum_r \dim [\bar{n}(r)] = 26.
\end{gather}
\end{subequations}
Here the expressions for the conformal-weight shifts
$\hat{\delta}_{0j}\ps$ are the same as those 
given in Eq.~(5.8), now slightly rearranged to emphasize the scaling into
the characteristic ratio
$(\fjs\frac{n(r)}{\rho(\sigma)})$ of the reduced formulation. The explicit form of the 
mode-normal ordering here
\begin{multline}
\nosub{
J_{n(r)\mu j} (M+\fjs\nrovrhos)
J_{n(s)\nu \ell}(N+\fls\nsovrhos)}{M}\\
=
\theta((M+\fjs\nrovrhos)\geq 0)
J_{n(s)\nu\ell}(N+\fls\nsovrhos)
J_{n(r)\mu j}(M+\fjs\nrovrhos)
\\
{}
+
\theta((M+\fjs\nrovrhos)< 0)
J_{n(r)\mu j}(M+\fjs\nrovrhos)
J_{n(s)\nu\ell}(N+\fls\nsovrhos)
\end{multline}
is also obtained as the image of the original mode-ordering in Eq.~(5.2). 
This result reflects the simple fact that the map (8.1) preserves the sign of the mode
number of each operator. 

As emphasized for the case of a single cycle of length two in Ref.~[3],
the reduced Virasoro generators (8.5) at $c_{j}(\sigma)=26$ are generically unconventional
in form: We remind that the spectral data of each element $\omega(\sigma) \in 
H'_{26}$ is recorded in the conventional orbifold fraction 
$n(r)/\rho(\sigma)$. The cycle length $f_{j}(\sigma)$ in the 
characteristic ratio $(\fjs\frac{n(r)}{\rho(\sigma)})$ seen here
represents the effect on $\omega(\sigma)$ due to the unwinding of cycle $j$ in each element of the basic
permutation group $H({\rm perm})_K$ of the orbifold-string theories.
Beyond the orbifold program and the reduction procedure described here, we are presently unaware of
any alternate path to these new Virasoro generators.

%MANU 8.6
Two further remarks are relevant before discussing the
reduced form of the extended physical-state
condition in the following section. The first remark concerns the \emph{target space-time structure}
of these theories, which is \emph{invariant} under
the reduction. In particular, each zero mode and hence the space-time 
dimension of each cycle is unchanged by the map
\begin{subequations}
\label{8.6.group1}
\begin{gather}
\label{8.6.1}
J_j (0)_\sigma = \hat{J}_j (0)_\sigma, \quad
\theta\{ J_j(0)_\sigma\} = 
\theta\{ \hat{J}_j(0)_\sigma\},
\\
\label{8.6.2}
D_j\ps = \hat{D}_j\ps
\end{gather}
\end{subequations}
and in fact the reduced formulation gives us a slightly more
uniform labeling of the momenta and the momentum-squared
operator of each cycle:
\begin{subequations}
\label{8.6.group2}
\begin{gather}
\label{8.6.3}
\{ J_j(0)_\sigma \}: \quad
J_{\frac{\rho(\sigma)\hj'}{\fjs},\mu(\frac{\rho(\sigma)\hj'}{\fjs}),j}(0),
\quad\hj' = 0,1,\dots,\fjs-1,
\\
\label{8.6.4}
\begin{aligned}[b]
P^2_j\ps& = \hat{P}^2_j\ps\\
{}
&= -\sum_{\mu,\nu}
\sum_{\hj'=0}^{\fjs-1}
\theta\{ J_j(0)_{ \sigma }\}
\mathcal{G}^{\frac{\rho(\sigma) \hj'}{\fjs},\mu; \frac{-\rho(\sigma)\hj'}{\fjs},\nu}\ps
J_{\frac{\rho(\sigma) \hj'}{\fjs},\mu j}(0)
J_{\frac{-\rho(\sigma) \hj'}{\fjs},\nu j}(0).
\end{aligned}
\end{gather}
\end{subequations}
%MANU 8.7
Here the type I zero modes are included at $\hat{j}'=0$ (see Eq.~(6.5b)), 
and the form of the momentum-squared operator  $\hat{P}^2_j\ps$ at $\hat{c}\ps = 26\fjs$ was given in 
Eq.~(6.7b). Similarly, the reduced number operator $R_{j}(\sigma)$ in the decomposition of $L_j(0)$
is the same
\begin{subequations}
\label{8.7.group2}
\begin{gather}
\label{8.7.1}
L_j(0) = \tfrac12 (-P^2_j\ps + R_j \ps) + \hat{\delta}_{0j}\ps,
\\
\label{8.7.2}
\begin{aligned}[b]
R_j\ps &= \hat{R}_j\ps\\
{} &= 
\Bigl(\sum_{n(r)\mu\nu}\sum_{P\in\Z}\Bigr)'
\nosub{
J_{n(r)\mu j}(P+\fjs\nrovrhos)
J_{-n(r),\nu j}(-P-\fjs\nrovrhos)
}{M}
\end{aligned}
\end{gather}
\end{subequations}
where the $\hat{c}\ps = 26\fjs$ form of
$\hat{R}_j\ps$ was given in Eq.~(7.4c).

%MANU 8.8
The second remark concerns the twist-field
state $\ket{0}_{j \sigma }$ whose definition in 
Eq.~(5.13) translates in the
$c_j\ps = 26$ formulation to 
\begin{equation}
\label{8.8.1}
J_{n(r)\mu j }((M+\fjs \nrovrhos)\geq 0)
\ket{0}_{j \sigma}=0.
\end{equation}
Under the action of the reduced Virasoro generators,
we find then that the conformal weights of this state are shifted 
as follows
\begin{subequations}
\label{8.8.group1}
\begin{gather}
\label{8.8.2}
(L_j(M\geq 0) - \kd{M}{0}\hat{\delta}_{0j}\ps ) 
\ket{0}_{j \sigma } = 0,\\
\label{8.8.3}
(L_\sigma(M\geq 0) - \kd{M}{0}\hat{\delta}_{ \sigma  })
\ket{0}_{\sigma } = 0,\\
\label{8.8.4}
\tsty
\ket{0}_{ \sigma } = 
\bigotimes_j \ket{0}_{j \sigma } , \quad
\hat{\delta}_{ \sigma } \equiv \sum_j \hat{\delta}_{0j}\ps
\end{gather}
\end{subequations}
where $L_{ \sigma }(M) = \sum_j L_j(M)$ are the total Virasoro
generators of sector $\sigma$. 

In summary so far, the  conformal-field-theoretic shifts
we have observed in the reduction 
\begin{subequations}
\label{8.8.group2}
\begin{gather}
\label{8.8.5}
\hat{c}_j\ps = 26\fjs \longrightarrow c_j \ps = 26,\\
\label{8.8.6}
\hat{\Delta}_{0j} \ps \longrightarrow \hat{\delta}_{0j}\ps,\\
\label{8.8.7}
j = 0,1,\dots,N\ps -1 ,\\
\label{8.8.8}
\hat{c}\ps = 26K \longrightarrow c\ps = 26 N(\sigma)
\end{gather}
\end{subequations}
are generalizations of the (inverse of) the central
charge and conformal-weight shifts found in
the original orbifold-induction procedure [7]  
and Ref.~[3].

\section{Equivalent $\cjs=26$ description of the physical states}\label{Section 9}
%MANU 9.1
There are two interpretations of the map (8.1) and its inverse. In the 
original conformal-field-theoretic interpretation of Ref.~[7], the results of the previous section
provide the construction (by relabeling) of one \emph {distinct} CFT in terms of another.
%By itself, the inverse orbifold-induction procedure (8.1)
%is only a relabeling of the operators of the orbifold-CFTs
%of permutation-type. The central part of this discussion is
We are not directly concerned with this CFT interpretation here.
There is however a second interpretation of the reduction procedure for the orbifold-\emph{string} theories of 
permutation-type, as restricted by the extended physical-state conditions 
(7.2) or (7.8) at $\hat{c}_j\ps = 26\fjs$.
In this string-theoretic interpretation, the map gives us a 
\emph {completely equivalent} $c_j\ps=26$ 
description of the physical spectrum of each cycle~$j$ in every sector $\sigma$ of 
the new string theories.
%originally described at $\hat{c}_j\ps = 26\fjs$ in the unreduced formulation.

Indeed, it is easily checked that all the 
components $\bhj = 0,1,\dots,\fjs-1$ of the extended 
physical-state conditions (7.2) of cycle~$j$ map directly onto the simpler physical-state 
condition of cycle~$j$ in the reduced $c_j\ps = 26$
description:
\begin{subequations}
\label{9.1.group1}
\begin{gather}
\label{9.1.1}
(L_j (M\geq 0) -\kd{M}{0})
\kets{\chi\ps}{j} = 0,\\
\label{9.1.2}
j = 0,1,\dots,N\ps-1.
\end{gather}
\end{subequations}
Here the reduced mode-ordered Virasoro generators $\{L_{j}(M)\}$ of cycle $j$ are given in 
Eq.~(8.5) and  $N(\sigma )$ is the number of cycles in sector~$\sigma$.
Note that the reduced physical state conditions (9.1) are 
\emph{conventional}, in that they exhibit \emph{unit intercept} for each 
cycle-string $j$. We finally emphasize with Ref.~[3] that the states described here
are \emph{exactly the same physical states} $\kets{\chi\ps}{j}$ -- now
%MANU 9.2 $\kets{\chi\ps}{j}$ -- now
rewritten in terms of the reduced currents --
which were originally defined by the extended physical-state conditions (7.2)
of the unreduced formulation at $\hat{c}_j\ps = 26\fjs$.

For example, assuming again that the zero-modes
$\{J_j(0)_{ \sigma } \} = \{\hat{J}_j(0)_{ \sigma}\}$ 
of cycle~$j$ in sector~$\sigma$ 
are
not an empty set, the physical ground-state of 
cycle~$j$ in sector~$\sigma$ is the \emph {same} momentum-boosted
twist-field state as determined earlier (see Eq.~(7.5)) in the
unreduced formulation:
\begin{subequations}
\label{9.2.group1}
\begin{gather}
\label{9.2.1}
\ket{0,J_j(0)}_{ \sigma } = \ket{0,\hat{J}_j(0)}_{ \sigma },\\
\label{9.2.2}
J_{n(r)\mu j}(( M+\fjs\nrovrhos) > 0 )
\ket{0,J_j(0)}_{\sigma} = 0,\\
\label{9.2.3}
L_{j}(M > 0)\ket{0,J_j(0)}_{ \sigma } = 0, \\
\label{9.2.4}
P^2_j\ps 
\ket{0,J_j(0)}_{\sigma} =
P^2_j\ps _{(0)}
\ket{0,J_j(0)}_{\sigma},\\
\label{9.2.5}
P^2_j\ps _{(0)} 
=
\hat{P}^2_j\ps _{(0)} 
= 2(\hat{\delta}_{0j}\ps - 1)\geq -2.
\end{gather}
\end{subequations}
Similarly, using the commutator (8.2b), the 
decomposition (8.9a) and the reduced physical-state conditions (9.1), one finds the level-spacing
in the reduced description of cycle~$j$ as
%MANU 9.3
\begin{equation}
\label{9.3.1}
\Delta(P^2_j\ps) = \Delta(R_j\ps) = 
2\abs{M+\fjs\nrovrhos}.
\end{equation}
This spacing results when a negatively-moded reduced
current $J_{n(r)\mu j} ((M+\fjs\nrovrhos)<0)$ is
added to any previous state $J...J\ket{0,J_j(0)}_{\sigma}$. 
%(Here we are using the
%map again to express the physical states in terms of the reduced currents.) 
Recalling that $M=\fjs m + \hj$ in the fundamental range of $\hj$, 
%\hj=0,...,f_{j}(\sigma)-1$,
these increments are recognized as the \emph{same} increments
(7.6) obtained in the $\hat{c}_j\ps = 26\fjs$ description of
the cycle.

%MANU 9.4
Finally a right-mover copy of the reduced physical-state conditions
at $c_{j}(\sigma)=26$ must be added 
%on the same physical states $\kets{\chi\ps}{j}$,
\begin{equation}
\label{9.4.1}
(L_j^L(M\geq 0) - \kd{M}{0})
\ket {\chi\ps}_j
 =
(L_j^R(M\geq 0) - \kd{M}{0})
\ket {\chi\ps}_j
=
0
\end{equation}
to obtain the reduced description of the physical states of
the closed-string sectors of the generalized permutation-orbifolds.
Again, these are the \emph {same} physical states described at $\hat{c}_j\ps = 26\fjs$
in Eq.~(7.8).
The reduced system (9.4) decomposes as follows
\begin{subequations}
\label{9.4.group1}
\begin{gather}
\label{9.4.2}
J_j^R(0)_{\sigma} = J_j^L(0)_{\sigma} 
=
\tfrac{1}{\sqrt2} J_j(0),\\
\label{9.4.3}
P_j^2\ps^R = P_j^2\ps^L =\tfrac12 P_j^2\ps,\\
\label{9.4.4}
P_j^2\ps\kets{\chi\ps}{j} = 2(P_j^2\ps_{(0)} +
R^L_j\ps )
\kets{\chi\ps}{j},\\
\label{9.4.5}
(R^R_j\ps - R^L_j\ps)
\kets{\chi\ps}{j} = 0
,\\
\label{9.4.6}
L^R_j(M>0) \kets{\chi\ps}{j}
=
L^L_j(M>0) \kets{\chi\ps}{j}
= 0,\\
\label{9.4.7}
P^2_j\ps^{\lower2pt\vbox{}\text{closed}}_{(0)} 
=
\hat{P}^2_j\ps^{\lower2pt\vbox{}\text{closed}}_{(0)} 
=
4(\hat{\delta}_{0j}\ps - 1)
\geq -4
\end{gather}
\end{subequations}
so that the \emph{same} value (7.11) of the ground-state momentum-squared
is obtained as well in the reduced formulation 
of the closed-string sectors.

\section{Example: The ``pure'' permutation orbifolds}
%MANU 10.1
The general formulae above are 
model-dependent, both in regard to $H({\rm perm})_{K}$ (where we have been 
explicit), and especially  on the choice
of element $\omega( \sigma)  \in H'_{26}$ (whose information is encoded in
the quantities $\{n(r)\mu, \mathcal{G}_{\cdot}\ps \}$).
These formulae will be used extensively in succeeding papers to study
large classes of models with explicit, non-trivial $H'_{26}$. 
%in succeeding papers of this series.

In this paper, however, we limit ourselves 
only to the very simplest orbifold-string theories
of permutation-type
\begin{equation}
\Bigl[\frac{{\rm U}(1) ^{26K}}{H({\rm perm})_K}\Bigr]_{\text{open}},
\quad
\Bigl[\frac{{\rm U}(1) ^{26K}}{H({\rm perm})_K}\Bigr]
\end{equation}
that is, the closed- and open-string analogues of the 
``pure'' permutation orbifolds with trivial $H'_{26}$.
The sectors $ \sigma $ of these orbifolds are described
by the equivalence classes of $H({\rm perm})_K$ alone, and
the open-string sectors of the orientation-orbifolds $U(1)^{26}/\Z_{2}(w.s.)$
are included in the open-string analogues when $K=2$.

%MANU 10.2
For these cases, the solutions to the $H'_{26}$ eigenvalue
problem (2.4a) is very simple:
\begin{subequations}
\label{10.2.group1}
\begin{gather}
\label{10.2.1}
\omega( \sigma)  = U\ps = 1,\quad \rho(\sigma)=1,\quad \bar{n} (0) = 0,\\
\label{10.2.2}
\mu = a = 0,1,\dots,25,\\
\label{10.2.3}
\mathcal{G}^{\cdot}\ps = \mathcal{G}_{\cdot}\ps = G_{ab} = -\eta_{ab},
\quad
\eta = \Bigl(
\hskip-2pt
\begin{array}{cr}
 1 & 0 \\
 %[-7pt]
 0 & - \thickone
 %\mathbb{1}
\end{array}
\hskip-2pt
\Bigr),
\\
\label{10.2.4}
\tsty
\sum_\mu = \sum_a = \dim [\bar{n}(0)] = 26.
\end{gather}
\end{subequations}
Here we have chosen the single spectral index in 
$\{\bar{n}(r)\}$ to be
$\bar{n}(0) = 0$, with degeneracy 26 described by the degeneracy index 
$\mu = a$. The 26-dimensional Minkowski metric $\eta$ is inherited directly
from the untwisted copies of the critical closed-string $U(1)^{26}$.

%MANU 10.3
With the data (10.3) for trivial $H'_{26}$, the orbifold Virasoro generators and the algebras
of twisted sector~$\sigma$ are easily read from
Eqs.~(4.4), (5.5b) and (5.8): 
\begin{subequations}
\label{10.3.group1}
\begin{gather}
\label{10.3.1}
\begin{gathered}[b]
\hat{L}_{\hj j} (m+\hatjovfs) = 
\kd{m+\hatjovfs}{0}
\hat{\Delta}_{0j}\ps
\hfill\\
\hskip7em{}
-\mfrac{1}{2\fjs}
\eta^{ab}
\sum_{\hat{\ell}=0}^{\fjs - 1}
\sum_{p\in \Z}
\nosub{
\hat{J}_{0a\hat{\ell} j}(p+\hatlovfjs)
\hat{J}_{0b,\hj-\hat{\ell},j}(m-p+\tfrac{\hj-\hat{\ell}}{\fjs})}{M},
\end{gathered}\\
\label{10.3.2}
\hat{\delta}_{0j}\ps = 0,\quad
\hat{\Delta}_{0j}\ps = \mfrac{13}{12}\Bigl(\fjs -\mfrac{1}{\fjs}\Bigr),
\hskip12em
\\
\label{10.3.3}
\begin{gathered}[b]
\hbox{}
[\hat{L}_{\hj j}(m+\hatjovfs),
\hat{L}_{\hj j}(n+\hatlovfs)]\hfill\\
\quad{}
=
\delta_{j\ell}
\bigl\{
(m-n-\tfrac{\hj-\hat{\ell}}{\fjs})
\hat{L}_{\hj + \hat{\ell},j}(m+n+\tfrac{\hj+\hat{\ell}}{\fjs})\\
\hskip 8em{}
+
\mfrac{26\fjs}{12}(m+\hatjovfs)
((m+\hatjovfs)^2 - 1)
\kd{m+n+\frac{\hj+\hat{\ell}}{\fjs}}{0}\bigr\},
\end{gathered}
\\
\label{10.3.4}
\begin{gathered}[b]
{}
[\hat{L}_{\hj j}(m+\hatjovfs), 
\hat{J}_{0a\hat{\ell}\ell}(n+\hatlovfs)]
%\hfill\\ \quad{}
= -\delta_{j\ell}(n+\hatlovfs)
\hat{J}_{0a,\hj+\hat{\ell},j}(m+n+\tfrac{\hj+\hat{\ell}}{\fjs}),
\end{gathered}\\
\label{10.3.5}
\begin{gathered}[b]
{}
[\hat{J}_{0a\hj j}(m+\hatjovfs), \hat{J}_{0b\hat{\ell} \ell}(n+\hatlovfs)]
%\hfill\\ \quad{}
= \delta_{j\ell} \eta_{ab}\fjs (n+\hatlovfjs)\kd{m+n+\frac{\hj+\hat{\ell}}{\fjs}}{0},
\end{gathered}\\
\label{10.3.6}
\begin{gathered}[b]
\tsty
\bhj = 0,1,\dots,\fjs-1,\quad
a = 0,1,\dots,25,\\
j = 0,1,\dots,N\ps - 1,\quad
\sum_j\fjs = K.
\end{gathered}
\end{gather}
\end{subequations}
We remind that $f_{j}(\sigma)$ is the length of cycle $j$ in sector 
$\sigma$ and the summation convention is assumed for
repeated indices $a,b$.
%MANU 10.4
To this list, we may add the periodicity conditions
\begin{subequations}
\label{10.4.group1}
\begin{gather}
\label{10.4.1}
\hat{L}_{\hj + \fjs, j}(m+\tfrac{\hj \pm \fjs}{\fjs}) =
\hat{L}_{\hj j}(m\pm 1 +\hatjovfs),\\
\label{10.4.2}
\hat{J}_{0a,\hj\pm\fjs,j}(m +\tfrac{\hj\pm\fjs}{\fjs}) = 
\hat{J}_{0a\hj j}(m\pm 1 +\hatjovfs)
\end{gather}
\end{subequations}
and the adjoint operations in these theories
\begin{subequations}
\label{10.4.group2}
\begin{gather}
\label{10.4.3}
\hat{J}_{0a\hj j}(m+\hatjovfs)^\dagger =
\hat{J}_{0a,-\hj,j}(-m-\hatjovfs),\\
\label{10.4.4}
\hat{L}_{\hj j}(m+\hatjovfs)^\dagger = 
\hat{L}_{-\hj,j}(-m-\hatjovfs), \quad
\hat{L}_{0j}(m)^\dagger = \hat{L}_{0j}(-m),\\
\label{10.4.5}
\begin{gathered}[b]
\norm{\hat{J}_{0a\hat{\ell}\ell}((m+\hatlovfs)< 0 )
\ket{0,\hat{J}_{l}(0)}_{ \sigma }}^2
%\hfill\\ \quad{}
 =
G_{aa}\fls\abs{m+\hatlovfs}\,
\bnorm{\ket{0,\hat{J}_{l}(0)}_{ \sigma }}^2.
\end{gathered}
\end{gather}
\end{subequations}
In fact, Ref.~[11] gives a form for the 
adjoint operation in any current-algebraic orbifold, but we give
this result here only for these simple cases.
Note in particular that the only negative-norm basis states
are associated here to the time-direction $a=0$ with
$G_{00}=-1$.

%MANU 10.5
We also remind that the cycle and sector central charges of these theories are
%MANU 10.5'
\begin{equation}
\label{10.5.1}
\hat{c}_j\ps = 26\fjs,\quad
\hat{c}\ps = \sum_j \hat{c}_j\ps = 26K.
\end{equation}
The system above is therefore an abelian limit
of the results given for the pure WZW permutation orbifolds
[15] with $\hat{c}_j\ps = c_g\fjs$ and $\hat{c} (\sigma)  = K c_g$, where
$c_g$ is the central charge of the affine--Sugawara construction  
[22] on $g$.

%MANU 10.6
The zero-modes (momenta) of cycle~$j$ in sector~$\sigma$ are
entirely of type I (see Sec.~6) in all these cases
\begin{subequations}
\label{10.6.group1}
\begin{gather}
\label{10.6.1}
\{\hat{J}_j(0)_{ \sigma }\} = 
\{ \hat{J}_{0a0j}(0), a=0,1,\dots,25\},\\
\label{10.6.2}
\hat{D}_j\ps = 26,\quad\tsty \hat{D}\ps = \sum_j \hat{D}_j\ps = 26 N\ps,\\
\label{10.6.3}
\hat{P}^2_j\ps = \eta^{ab}\hat{J}_{0a0j}(0)\hat{J}_{0b0j}(0),\\
\label{10.6.4}
\hat{P}^2_j\ps_{(0)} = -2,\quad
\Delta (\hat{P}_j^2\ps) = 2\fjs \abs{m+\hatjovfs}
\end{gather}
\end{subequations}
where $\hat{D}_j\ps = 26$ is the target space-time dimension
of cycle~$j$ in sector~$\sigma$ and $N\ps$ is 
the number of cycles in sector~$\sigma$. We
note in particular that each cycle~$j$, though described
here at $\hat{c}_j\ps = 26\fjs$, has exactly 26 space-time dimensions
at ground-state mass-squared~$-2$, just as in an
ordinary untwisted critical open string. Of course,
the data given so far is for the open-string sectors
of $[{\rm U}(1)^{26K} / H({\rm perm})_K]_{\text{open}}$, whereas a right-mover
copy is needed to describe the closed-string
sectors of ${\rm U}(1)^{26K}/H({\rm perm})_K$. In the latter cases we 
find that each cycle has
26 left- and right-mover momenta and $\hat{P}^2_j\ps^{\text{closed}} = -4$,
again the same as an ordinary untwisted critical closed string.
The open- and closed-string forms of the extended physical-state conditions
%MANU 10.6'
%extended physical state conditions for the open- and closed-string
%analogues of the pure permutation orbifolds are given
are given respectively in Eqs.~(7.2) and~(7.8).

%MANU 10.7
Let us turn finally to the equivalent, reduced formulation
of these theories at reduced cycle central charge $c_j\ps = 26$, where one finds
from Eqs.~(8.2),(8.5) and (10.2) that
\begin{subequations}
\label{10.7.group1}
\begin{gather}
\label{10.7.1}
L_j(M) = -\tfrac12 \eta^{ab}\sum_{P\in \Z}
\nosub{
J_{0aj}(P)
J_{0bj}(M-P)}{M},\\
\label{10.7.2}
[L_j(M), L_{\ell}(N)] =
\delta_{j\ell}\{(M-N)L_j(M+N)+\tfrac{26}{12}M(M^2-1)
\kd{M+N}{0}\},\\
\label{10.7.3}
[L_j(M),J_{0a\ell}(N)] = -\delta_{j\ell}N J_{0a\ell}(M+N),\\
\label{10.7.4}
[J_{0aj}(M),J_{0b\ell}(N)] = -\delta_{j\ell}\eta_{ab}N\kd{M+N}{0},\\
\label{10.7.5}
a=0,1,\dots,25,\quad j = 0,1,\dots,N\ps -1.
\end{gather}
\end{subequations}
As discussed more generally in Sec.~8, the zero-modes (momenta) in the reduced formulation
are isomorphic to those in the unreduced
formulation:
\begin{subequations}
\label{10.7.group2}
\begin{gather}
\label{10.7.6}
\{J_j(0)_{ \sigma }\} = 
\{\hat{J}_j(0)_{ \sigma }\} = 
\{J_{0aj}(0),a=0,1,\dots,25\},\\
\label{10.7.7}
D_j\ps = \hat{D}_j\ps = 26= c_j\ps,\quad
D\ps=\hat{D}\ps = 26 N\ps = c\ps,\\
\label{10.7.8}
P^2_j\ps =
\hat{P}^2_j\ps =\eta^{ab}J_{0a0j}(0)J_{0b0j}(0),\\
\label{10.7.9}
P^2_j\ps_{(0)} = 
\hat{P}^2_j\ps_{(0)} = -2,\quad
\Delta
P^2_j\ps_{(0)} = 
\Delta
\hat{P}^2_j\ps_{(0)} = 2\mskip1mu\abs{M}.
\end{gather}
\end{subequations}
The last result is the level-spacing induced
by adding an
extra negatively-moded current $J_{0aj}(M<0)$ to a lower-level state and,
recalling that $M_j=m\fjs+\bhj$, 
one checks that this level-spacing is indeed the same
as that given in Eq.~(10.7d) for the unreduced currents.

%MANU 10.8
We end our technical discussion with the adjoints and norms in the reduced
formulation 
\begin{subequations}
\label{10.8.group1}
\begin{gather}
\label{10.8.1}
J_{0aj}(M)^\dagger = J_{0aj}(-M),\quad
L_j(M)^\dagger = L_j(-M),\\
\label{10.8.2}
\bnorm{J_{0a\ell}(M < 0) \ket{0,J_l(0)_{ \sigma }}}^2
= G_{aa}\abs{M}\,
\bnorm{\ket{0,J_l(0)_\sigma}}^2
\end{gather}
\end{subequations}
where the adjoints are the map of Eqs.~(10.5a,b) and the norms are 
computed from the adjoints. Again using $M_{\ell} = \fls m+\bar{\hat{\ell}}$, we see that the norms
are the same as those computed in Eq.~(10.5c) for the original 
formulation. Similarly, of course, the inner product of any two states are the same in the reduced and
unreduced formulations.

%MANU 10.9
Taken together with the reduced physical-state conditions (9.1) 
and (9.4) for the open- and closed-string sectors, these results allow us to
conclude the following on inspection: 
%for the ``pure'' orbifold-string theories of permutation-type, 
Each cycle~$j$ of each sector~$\sigma$ of the ``pure'' orbifold-strings (10.1)
is nothing but an ordinary untwisted 26-dimensional string with
target space-time symmetry $\operatorname{SO}(25,1)$. This conclusion is however
quite special for the ``pure'' orbifold-string systems with trivial 
$H'_{26}$, whereas (as seen for $H({\rm perm})_2=\Z_2$ and
$\Z_2({\rm w.s.})$  in Ref.~[3]) the orbifold-string theories with
non-trivial $H'_{26}$ are generically new.

%MANU 10.9
The critical-string equivalences of this section were anticipated for the ``pure'' orbifolds of
permutation-type with $H({\rm perm})_2=\Z_2$ or
$\Z_2({\rm w.s.})$ in Ref.~[3], and
were verified at the interacting level [5] for the pure permutation
orbifolds with $H({\rm perm})_K=\Z_K$, $K$ prime. Moreover,
our conclusion here was 
conjectured for all $H({\rm perm})_K$
in Ref.~[5]. It should be added that special cases with particular non-trivial $H'_{26}$ (see 
e.g. the orientation-orbifold string system in Ref.~[4]) can also 
be equivalent to ordinary critical strings, including the critical bosonic
open-closed string system. Taken together then, the orbifold-string systems of
permutation-type provide several rising, ever-more twisted hierarchies of 
new string theories,
including ordinary critical strings as the simplest cases.

%MANU 10.10
We finally note that the pure permutation orbifolds
\begin{equation}
\label{10.10.1}
 \frac{{\rm U}(1)^{26K}}{H({\rm perm})_K},
\end{equation}
being composed entirely of ordinary closed-string
sectors, exhibit \emph{multiple gravitons}. Indeed, we have seen here that the free theories
in these cases exhibit one graviton per cycle per sector. The only
examples studied so far at the symmetrized, interacting level are the prime cyclic
permutation orbifolds
\begin{equation}
\label{10.10.2}
\frac{{\rm U}(1)^{26\lambda}}{\Z_\lambda}, \quad\hbox{$\lambda$ prime}
\end{equation}
where the linear (diagonal) modular-invariant
construction of Ref.~[5] shows the total number of
gravitons
\begin{equation}
\label{10.10.3}
N_\lambda = \lambda+(\lambda-1) = 2\lambda-1.
\end{equation}
This includes in particular one graviton in each nontrivial
twisted sector. The interaction (or presumably non-interaction)
among these gravitons will require the construction
of the twist fields (intertwiners) among the sectors, an inquiry which
%MANU 10.11
is beyond the scope of this paper. We similarly
expect more than one graviton in the generalized
permutation orbifolds with nontrivial
$H'_{26}$, at least from the cycles of the sector corresponding to the
unit elements of $H({\rm perm})_K$ and $H'_{26}$. On the other
hand -- as we will discuss in succeeding papers -- the open-closed string systems of the orientation-orbifolds [3,4]
\begin{equation}
\label{10.10.4}
\frac{{\rm U}(1)^{26 K}}{H_-} = \frac{{\rm U}(1)_L^{26}\times{\rm U}(1)_R^{26}}{H_-},
\quad
H_{-}\subset \Z_2({\rm w.s.}) \times H'_{26}
\end{equation}
have only a single graviton for any choice of $H'_{26}$.

\section{Conclusions}
%MANU 11.1
In the previous paper [6] of this series, we used BRST quantization to 
find the  orbifold Virasoro algebras and extended physical-state conditions of the bosonic prototypes
% of
%To supplement the extended physical-state
%conditions of the previous paper [6] in this series,
%we have provided the general form of
%the orbifold Virasoro generators
of the orbifold-string theories of permutation-type:
\begin{subequations}
\label{11.1.group1}
\begin{gather}
\label{11.1.1}
\frac{{\rm U}(1)^{26K}}{H_+},\quad
\Bigl[\frac{{\rm U}(1)^{26K}}{H_+}\Bigr]_{\text{open}},\quad
H_+\subset H({\rm perm})_K\times H'_{26}\\
\label{11.1.2}
\frac{{\rm U}(1)^{26}}{H_-} = 
\frac{{\rm U}(1)^{26}_L\times U(1)^{26}_R}{H_-},\quad
H_-\subset \Z_2({\rm w.s.})\times H'_{26}.
\end{gather}
\end{subequations}
These theories live at cycle central charge
$\hat{c}_j\ps = 26\fjs$, where $\fjs$ is the
length of cycle~$j$ in each equivalence
class $ \sigma $ of the permutation group
$H({\rm perm})_{K}$ or $\Z_{2}(w.s.)$. The expected sector central charges 
$\hat{c}(\sigma)=26K$ of each orbifold are obtained by summing over the cycles of sector $\sigma$.

In this paper we have completed the cycle dynamics of these theories,
supplementing the extended physical-state conditions of cycle $j$ with the 
explicit form of the orbifold Virasoro generators as functions of the twisted
matter of each cycle. Our results here are general, depending on the choice
of element $\omega( \sigma)  \in H'_{26}$ in the divisors of each orbifold.
With these tools, we also began a systematic inquiry into the target space-time
structure of these theories, including in particular the number $\hat{D}_j\ps$ 
of target space-time dimensions in cycle $j$ of sector $\sigma$.
%An equivalent, 
%reduced formulation was also given for each cycle
%at $c_j\ps = 26$. At both levels, we gave 
%general formulae depending on
%the choice of element $\omega( \sigma)  \in H'_{26}$, emphasizing
%general formulae for the number of space-time dimensions
%$D_j\ps$ of the target space of cycle~$j$ in sector~$\sigma$.

We also found an equivalent, reduced description of the physical states
of each cycle at reduced cycle central charge $ c_j\ps = 26$, emphasizing
that the target space-time properties of the theories are invariant under the
reduction and in fact more transparent in the reduced formulation. 

%MANU 11.2
As examples, the simplest cases with trivial $H'_{26}$ (the  
orbifolds of ``pure'' permutation-type)
\begin{equation}
\frac{{\rm U}(1)^{26K}}{H({\rm perm})_K},
\quad
\Bigl[\frac{{\rm U}(1)^{26K}}{H({\rm perm})_K}\Bigr]_{\text{open}},\quad
\frac{{\rm U}(1)^{26}}{\Z_2({\rm w.s.})}
\end{equation}
were worked out in some detail, with the result that
$\hat{D}_j\ps = 26$ for each cycle~$j$ in each sector~$\sigma$ of 
these examples. Indeed, the reduced formulation transparently shows
that each of these cycles is spectrally equivalent to
an ordinary untwisted 26-dimensional string.

This is not the case however for the more general situation with
nontrivial $H'_{26}$, which provides large classes of new 
string theories. In particular, it is clear from our discussion
that the target space-time dimensionality of these theories is not 
generically equal to any of the central charges discussed above.
%which we will further discuss

In the following paper, we will apply the general formulae
developed here to study a large example of non-trivial $H'_{26}$, finding that
the new string theories in fact exhibit many target space-times of varying
dimensionality, symmetry and signature -- including in particular Lorentzian target space-times
with $\hat{D}_j\ps\leq 26$.

\section*{Acknowledgments}
For helpful discussions and encouragement, I thank
L.~Alvarez-Gaum\'e, C.~Bachas, J.~de Boer, 
S.~Frolov, O.~Ganor, E.~Kiritsis, A.~Neveu, H.~Nicolai,
N.~Obers, B.~Pioline, M.~Porrati, E.~Rabinovici, 
V.~Schomerus, C.~Schweigert, M.~Staudacher, R.~Stora,
C.~Thorn, E.~Verlinde and J.-B.~Zuber.

\clearpage

%\bibliographystyle{annals-alpha}
%\bibliography{OstII}

\providecommand{\bysame}{\leavevmode\hbox to3em{\hrulefill}\thinspace}
\providecommand{\xxMR}[1]{\relax\ifhmode\unskip\space\fi
  \href{http://www.ams.org/mathscinet-getitem?mr=#1}{MR~#1}}
\providecommand{\xxZBL}[1]{\relax\ifhmode\unskip\space\fi
  \href{http://www.emis.de/cgi-bin/MATH-item?#1}{ZBL~#1}}
\providecommand{\xxJFM}[1]{\relax\ifhmode\unskip\space\fi
  \href{http://www.emis.de/cgi-bin/JFM-item?#1}{JFM~#1}}
\providecommand{\xxARXIV}[1]{\relax\ifhmode\unskip\space\fi
  \href{http://arxiv.org/abs/#1}{arXiv~#1}}
\providecommand\bibmarginpar{\leavevmode\marginpar}
\providecommand{\href}[2]{#2}

\end{document}